\documentclass[12pt,preprint]{aastex}
\usepackage{epsfig}

\def\i{\hbox{\,{\sc i}}}
\def\ii{\hbox{\,{\sc ii}}}
\def\iii{\hbox{\,{\sc iii}}}
\def\iv{\hbox{\,{\sc iv}}}
\newcommand{\teff}{\mbox{$T_{\mathrm{eff}}$}}
\newcommand{\logg}{\mbox{$\log g$}}

\shorttitle{The Nature of Algol Disks}
\shortauthors{Miller et al.}

\begin{document}

\title{Revealing the Nature of Algol Disks through Optical and UV
Spectroscopy, Synthetic Spectra, and Tomography of TT Hydrae}

\author{Brendan Miller, J\'{a}n Budaj\altaffilmark{1,3}, Mercedes Richards\altaffilmark{2},} 

\affil{Department of Astronomy \& Astrophysics, Pennsylvania State University,
525 Davey Laboratory, University Park, PA, 16802, USA}
\email{bmiller@astro.psu.edu, budaj@as.arizona.edu, mrichards@astro.psu.edu} 

\author{Pavel Koubsk{\'y}\altaffilmark{2}}

\affil{Astronomical Institute, Academy of Sciences  of the Czech
Republic, 251\,65~Ond{\v r}ejov, Czech Republic}
\email{koubsky@sunstel.asu.cas.cz}

\and

\author{Geraldine J. Peters\altaffilmark{2}}

\affil{Space Sciences center, University of Southern California,
Los Angeles, CA 90089-1341}
\email{gjpeters@mucen.usc.edu}

\altaffiltext{1}{Astronomical Institute, Tatransk\'{a} Lomnica, 05960, 
Slovak Republic}
\altaffiltext{2}{Visiting Astronomer, Kitt Peak National Observatory, 
National Optical Astronomy Observatories.  NOAO is operated by AURA, Inc.
under contract to the NSF.} 
\altaffiltext{3}{New address: Department of Astronomy, University of Arizona,
933 N. Cherry Ave., Tucson, AZ 85721, USA}

\begin{abstract}

We have developed a systematic procedure to study the disks in Algol-type
binaries using spectroscopic analysis, synthetic spectra, and tomography.  We
analyzed 119 H$\alpha$ spectra of TT Hya, an Algol-type eclipsing interacting
binary, collected from 1985-2001.  The new radial velocities enabled us to
derive reliable orbital elements, including a small non-zero eccentricity,
and to improve the accuracy of the absolute dimensions of the system.  High
resolution IUE spectra were also analyzed to study the formation of the
ultraviolet lines and continuum.  Synthetic spectra of the iron curtain using
our new {\sc{shellspec}} program enabled us to derive a characteristic disk
temperature of 7000K.  We have demonstrated that the UV emission lines seen
during total primary eclipse cannot originate from the accretion disk, but
most likely arise from a hotter disk-stream interaction region.

The synthetic spectra of the stars, disk, and stream allowed us to derive a
mass transfer rate $\ge2\times10^{-10}M_{\odot}yr^{-1}$.  Doppler tomography
of the observed H$\alpha$ profiles revealed a distinct accretion disk.  The
difference spectra produced by subtracting the synthetic spectra of the stars
resulted in an image of the disk, which virtually disappeared once the
composite synthetic spectra of the stars and disk were used to calculate the
difference spectra.  An intensity enhancement of the resulting tomogram
revealed images of the gas stream and an emission arc.  We successfully
modeled the gas stream using {\sc{shellspec}} and associated the emission arc
with an asymmetry in the accretion disk.

\end{abstract}


\section{Introduction} 

The Algol-type binaries are interacting systems consisting of a hot,
usually more massive, main sequence star and a cool giant or subgiant
which fills its Roche lobe.  While these binaries should be randomly 
oriented, more eclipsing systems have been discovered since these 
systems are more easily detected through their distinct light curves.
By convention, the hotter star (associated with the deeper minimum) is
referred to as the ``primary'' and the cooler star as the ``secondary.''
The secondary is losing mass to its companion and we observe circumstellar
material in the form of a gas stream, an accretion annulus, or an
accretion disk.  The most recent paper that summarizes our current
knowledge of the Algols is the work by \citet{richards+albright99}.  Some
recent progress in the field includes:  the detection and study of the
pulsation of primary stars \citep{lehmann+mkrtichian04, mkrtichianetal05};
evidence of the superhump phenomenon in the radio power spectrum of
$\beta$ Per \citep*{retter+richards+wu05}; the calculation and modeling of
synthetic spectra \citep*{baraietal04, budaj+richards+miller05,
korcakovaetal05}; far UV and X-ray studies \citep*{chungetal04,
peters+polidan04}; Doppler tomography of the accretion structures
\citep{richards04, baraietal04}; eclipse mapping \citep*{pavlovskietal06};
new orbits and fundamental parameters of eclipsing systems \citep*{mmmwzkt05};
hydrodynamic simulations of the mass transfer process
\citep{richards+ratliff98, nazarenkoetal05}; a survey of H$\alpha$ mass
transfer structures \citep*{vesperetal01}; and a study of the evolution of
Algol systems \citep{maxted+hilditch96}.

TT Hydrae (HD97528, HIP54807, SAO179648, $V=7.27^{m}, \alpha=11^{h}13^{m},
\delta=-26^{\circ}28'$) is a long-period ($P_{\rm orb}=6.95$ days)
Algol-type eclipsing binary system which consists of a hotter B9.5 V main
sequence primary and a cooler evolved K1 III-IV secondary star which fills
its Roche lobe.  \citet{plavec+polidan76} found evidence of mass transfer
in the form of an accretion disk surrounding the primary.
\citet{kulkarni+abhyankar80} obtained UBV photometric observations and
elements of the system.  These observations were re-examined by
\citet{etzel88} who found from photometry, spectrophotometry, and
spectroscopy that $T_{\rm eff}=9800$K, $v\sin i=168\pm5$ km\,s$^{-1}$ for
the primary, and $T_{\rm eff}=4670-4850$K for the secondary, photometric
mass ratio $q=0.184$, and $i=84.415\pm0.042$.  He also estimated a
distance of 193 pc to the binary.  \citet{eaton+henry92} estimated $v\sin
i=43\pm3$ km\,s$^{-1}$ for the secondary and suggested that the mass
ratio, as well as the mass of the primary, should be slightly higher than
those found by \citet{etzel88} if the secondary fills its Roche lobe.

\citet{plavec88} studied the IUE spectra and concluded from the presence
of Fe\ii\ absorption lines that the vertical dimension of the accretion
disk is significant and is at least comparable to the diameter of the
primary.  He estimated a distance to the system of 194 pc and pointed out
that the presence of super-ionized emission lines of Si\iv, C\iv, and N{\sc v}
poses a question about the ionization source.  \citet{peters89}
estimated from the eclipses of the H$\alpha$ emission region that a fairly
symmetrical disk fills up to 95\% of the Roche lobe radius.  Moreover, she
argued that the disk must be rather flat with the vertical dimension
comparable to or less than the diameter of the primary.  The depth of the
H$\alpha$ core was found to be strongly variable with phase and was
deepest shortly before and after the primary eclipse, a feature also
observed in other Algols ({\it{e.g.}}, \citealt{richards93}).

\citet{sahade+cesco46} determined the orbital parameters of the primary
and identified a highly eccentric orbit from the Ca\ii~K and H lines.
\citet{miller+mcnamara63} measured a few radial velocities of the primary
and secondary.  \citet{popper89} measured radial velocities of the primary
and secondary and found that the radial velocities of the secondary were
consistent with a circular orbit ($K=132$ km\,s$^{-1}$) but this was not
true in the case of the primary.  The radial velocities of the primary
suggested a mass ratio of about $q=0.26$, which is not in very good
agreement with the photometric value.  \citet{vanhamme+wilson93}
reanalyzed the light and velocity curves separately and simultaneously and
used a physical model to obtain the following parameters:  separation
$a=22.63\pm0.12 R_{\odot}$, effective temperature of the primary
$T_{1}=9800$K, secondary $T_{2}=4361$K, radius of the primary $R_{1}=1.95
R_{\odot}$, secondary $R_{2}=5.87 R_{\odot}$, mass of the primary
$M_{1}=2.63M_{\odot}$, secondary $M_{2}=0.59M_{\odot}$, and mass ratio
$q=0.2261\pm0.0008$.  \citet{vivekananda+sarma94} reanalyzed the original
\citet{kulkarni+abhyankar80} data using the Wilson-Devinney code
\citep{wilson+devinney71} and obtained a photometric mass ratio of $q$ =
0.2963, which is very close to the spectroscopic value derived by
\citet{popper89}.

\citet{albright+richards96} obtained the first Doppler tomogram of TT Hya
and produced an indirect image of the accretion disk.
\citet{peters+polidan98} observed redshifted absorption in N\i, N\ii\
lines in their FUV spectra at the phase 0.95 and interpreted it as
evidence of the gas stream; consequently they were able to infer that the
rate of inflow was greater than $10^{-12}M_{\odot}yr^{-1}$.
\citet{richards+albright99} studied the H$\alpha$ difference profiles and
properties of the accretion region and placed TT Hya in the context of
other Algol binaries.  Most recently, \citet{budaj+richards+miller05}
produced synthetic H$\alpha$ spectra, compared the synthetic and observed
profiles, determined the physical properties of the disk, and found
evidence of the gas stream in these profiles.  They also suggested the
presence of an additional circumstellar structure between the C1 and C2
Lagrangian surfaces, in the vicinity of the secondary star, to explain the
strong H$\alpha$ line cores and other effects.

In this work, we have re-examined the earlier conclusions by analyzing
previously unpublished H$\alpha$ spectra of TT Hya.  These observations
are described in \S 2.  In \S 3, we describe the study of the optical
spectra, the redetermination of orbital elements and absolute dimensions
of the system from new velocity curves, and the study of the radial
velocities of the profiles.  In addition, synthetic H$\alpha$ spectra were
calculated, and a variety of difference spectra were determined by
subtracting synthetic spectra from the observed H$\alpha$ profiles.  In \S
4, archival high resolution ultraviolet spectra were used to provide a
consistent model for the accretion structures in TT Hya.  In \S 5, we show
how Doppler tomography was used to study the accretion structures in the
binary and as a tool to examine the quality of the fits between the
observed and synthetic spectra.  Finally, the conclusions are presented in
\S6.

\section{The Observations}
\label{s2}

The 119 optical spectra described in this work were obtained by various 
observers from 1985 to 2001 (see Table \ref{t1}).  All spectra were
collected with the 0.9m Coud{\'e} Feed Telescope at Kitt Peak National
Observatory (KPNO).  These spectra have a resolution of 0.34 \AA\ and
dispersion at H$\alpha$ of $\sim$7.0 \AA{~}mm$^{-1}$.  The data can be
separated roughly into two major groups:  1985 Feb -- 1991 Mar (Group 1)
and 1994 Apr - 2001 Jan (Group 2).  In addition, we used the original 27
radial velocity measurements of \citet{popper89} based on spectra
collected from 1956 - 1977 at Mt.  Wilson and Lick Observatories at a
resolution of 11--20 \AA/mm (see Table \ref{t1}).  The UT dates, HJD, and
epochs of the KPNO data are listed in the first three columns of Table
\ref{t2}.

The ultraviolet spectra of TT Hya were obtained with the IUE telescope
from 1980 to 1992.  The short wavelength range (1150\AA\ -- 1950\AA) was
covered by the SWP camera while the long wavelength range (1900\AA\ --
3200\AA) was covered by the LWP and LWR cameras \citep{kondo87}.  The high
resolution IUE spectra have a resolution of 0.1\AA\ to 0.3\AA\ or R=10000.
Only the high resolution spectra are studied here since the low resolution
spectra are not suitable for radial velocity studies.  Moreover, these
high resolution IUE spectra are similar in resolution to the optical
spectra.  Six long wavelength (LWP and LWR) spectra and eleven short
wavelength (SWP) spectra were analyzed.  These data are summarized in
Table \ref{t3}.

\begin{figure*}
\figurenum{1}
\epsscale{1.0}
\plotone{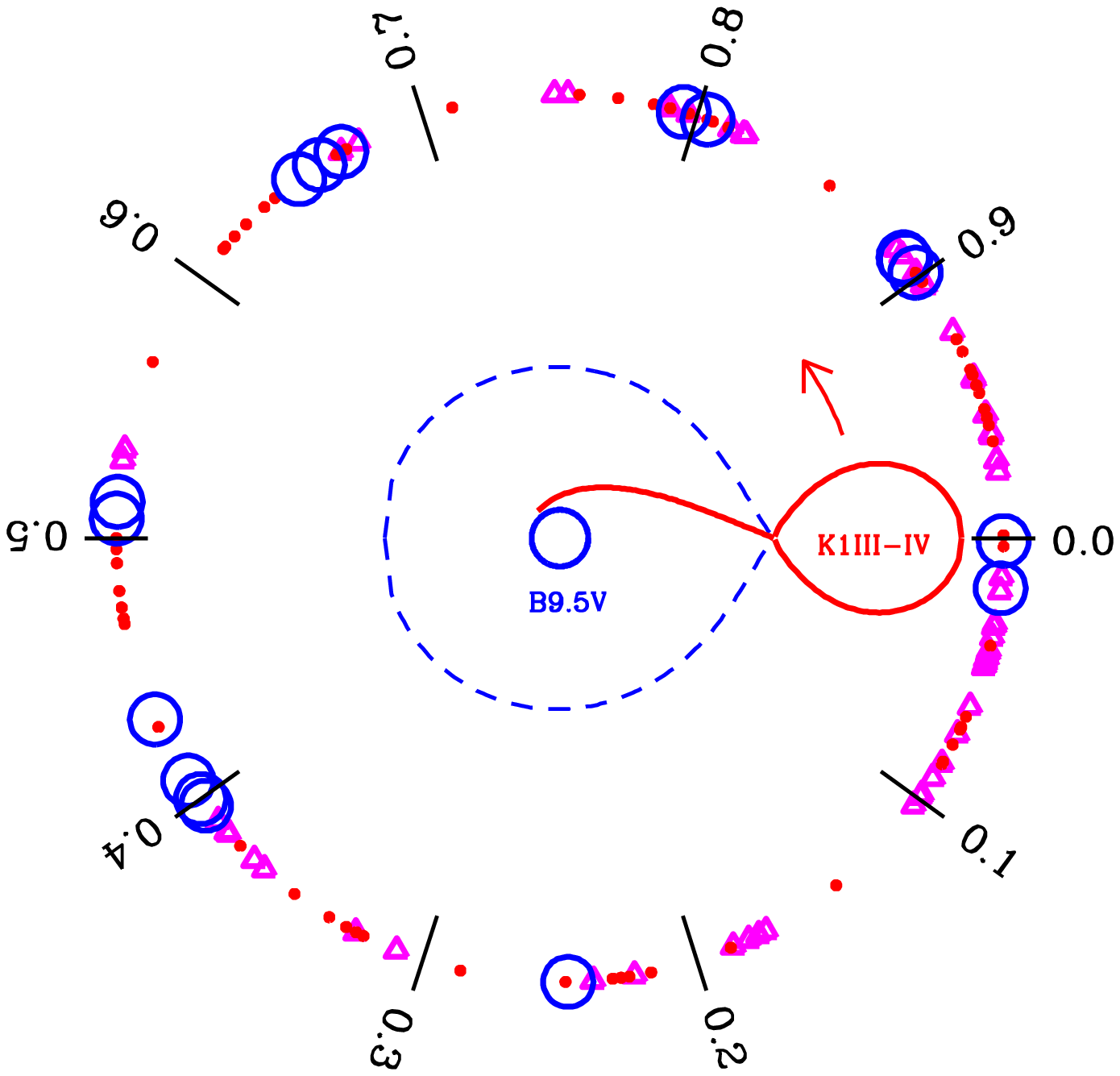}
\caption{Phase coverage of TT Hya showing the KPNO data collected from
1985 Feb --1991 Mar (open triangles) and 1994 Apr -- 2001 (solid circles),
as well as the high resolution IUE data collected from 1980 Dec to 1992
Dec (large open circles).}
\label{f1}
\end{figure*}

The photometric phases, $\phi$, were calculated by assuming the
photometric period, $P = 6.95342913$ \citep{kulkarni+abhyankar80} and the
ephemeris of primary minimum, $HJD_{min} = 2424615.3835$ \citep{etzel88}.
\begin{equation}
Epoch, E = (HJD - 2424615.3835)/6.95342913
\label{e1}
\end{equation}
In Tables 2 and 3, HJD is the Heliocentric Julian date, and Epoch.phase is
the whole number epoch with the photometric phase, $\phi$, listed after
the decimal point.  The phase coverage of the spectra is illustrated in
Figure \ref{f1}.

\section{The Optical Spectra}
\label{s3}

This section describes the analysis of the optical spectra.  In \S
3.1, we derive precise orbital elements and improve the values for the
absolute dimensions of the system based on a new velocity curve for the
secondary.  In \S3.2, we discuss the central absorption of the
H$\alpha$ line which is most closely associated with the primary star.  In
\S3.3, we describe the calculation of synthetic H$\alpha$ spectra with the new code
called {\sc{shellspec}} \citep{budaj+richards04, budaj+richards+miller05}.
Model spectra were calculated for the stars, accretion disk, and gas
stream.  These synthetic spectra were compared with the observed spectra
and both were later used to make reconstructed images of the accretion
structures in the binary (see \S5).

\subsection{Radial Velocity Measurements of the Secondary}

The spectral lines of the secondary are seen more clearly in our optical
spectra because they are much sharper than the lines of the primary.  They
are more accurate tracers of the orbital motion than the lines of the
primary, which are highly broadened by the stellar rotation and distorted
by the presence of the accretion disk.  Hence, orbital elements based on
the lines of the primary are not as precise as those derived from the
lines of the secondary.

The radial velocities were determined using the cross-correlation
technique and a template based on synthetic spectra was created for this
procedure.  First, the model atmosphere of the secondary star was obtained
by interpolating in the \citet{kurucz93a} \teff-\logg\ grid and assuming
solar abundances and a microturbulence of 2 km\,s$^{-1}$.  Next, the
intrinsic synthetic stellar spectrum of the secondary emerging from this
atmospheric model (flux per unit surface area) was calculated using the
code called {\sc{synspec}} \citep{hubenyetal94}, as modified by
\citet{krticka98}.  Here again, solar abundances and a microturbulence of
2 km\,s$^{-1}$ were assumed.  The synthetic spectrum was then convolved
with the rotational ($v\sin i=43 $ km\,s$^{-1}$; \citealt{eaton+henry92})
and instrumental profiles.  Subsequently, this synthetic spectrum was
diluted with the featureless continuum of the primary assuming a known
L1/L2 ratio, where L1 and L2 are the luminosities corresponding to the
H$\alpha$ continuum of the primary and secondary, respectively.  This
diluted spectrum was then used as a template for the radial velocity
measurements.  The radial velocities of the secondary star were measured
using the cross-correlation of the observed spectra with this template
synthetic spectrum, omitting the region near H$\alpha$.  Six spectra at
phases 3092.942, 3096.962, 3097.371, 3669.771, 3669.784, and 3737.902 were
excluded from the measurements due to either a low S/N or heavy blending
by terrestrial lines, hence a total of 113 highly reliable radial velocity
measurements of the secondary star were obtained.  These are listed under
$V_r(2)$ in column 8 of Table \ref{t2}.  Our measurements have more than
quadrupled the previous number of radial velocity measurements of the
secondary since Popper's work.  Moreover, the much improved precision and
phase coverage allowed us to examine whether Algol-type binaries might
have non-circular orbits.

The derivation of orbital elements for TT Hya was performed using the
{\sc{sbcm}} code of \citet{morbey+brosterhus74}.  Table \ref{t4} lists 
the resulting orbital elements: 
systemic velocity, $V_{0}$; velocity semi-amplitude, $K_{2}$; 
orbital eccentricity, $e$; longitude of periastron, $\omega$;
epoch of the periastron, $T_{0}$; orbital period, $P$; mass function, $f(m)$;
semi-major axis,  $a_{2}\sin i$; and standard deviation of the fit, $\sigma$.
First, the original 27 radial
velocity measurements of \citet{popper89} from 1956 -- 1977 were
reanalyzed and a mean standard deviation, $\sigma \sim 5.1$ km\,s$^{-1}$
was obtained.  Next, we analyzed our KPNO data in three groups by date:
1985--1991, 1994-2001, and 1985--2001 and found solutions with $\sigma =
4.4$ km\,s$^{-1}$, $\sigma = 2.0$ km\,s$^{-1}$, and $\sigma = 2.9$
km\,s$^{-1}$, respectively.  Finally, we combined our data with those of
\citet{popper89} to obtain a data set of 140 measurements over an extended
baseline of 45 years.  The measurements were weighted according to the
inverse square of their standard deviation, which resulted in weights of
0.15 for the 1956--1977 data from \citet{popper89}, 0.21 for the 1985-1991
data and 1.0 for the 1994-2001 data.  Since the precision of the
photometric period of the \citet{kulkarni+abhyankar80} was uncertain, we
calculated four velocity curve solutions.  The orbital period was fixed at
the value published by \citet{kulkarni+abhyankar80} for the analysis of
the (1) 1994--2001 data, (2) 1985--2001 data, and (3) the complete dataset
from 1956--2001 which included Popper's measurements (see Table \ref{t4},
columns 2 -- 4).  In the final case, the 1956--2001 data were analyzed
with the period as a free parameter within {\sc{sbcm}} (Table \ref{t4},
column 5).  The elements derived from column 5 of Table \ref{t4} were used
to calculate the (O-C) and spectroscopic phases listed in Table \ref{t2}.

\begin{figure*}
\figurenum{2}
\epsscale{1.0}
\includegraphics[angle=-90,scale=0.6]{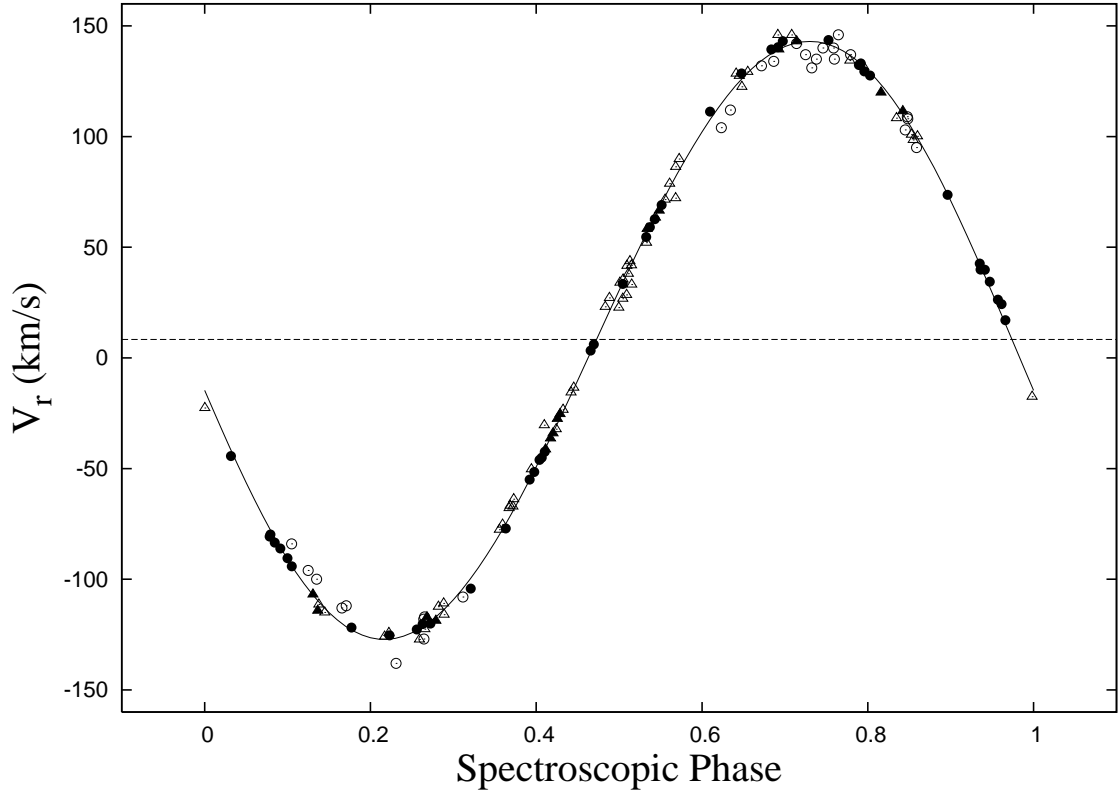}
\caption{
Velocity curve of TT Hya based on elements from Table \ref{t4}, column 5,
showing the TT Hya spectra obtained from 1985 Feb --1991 Mar (Peters -
open triangles), 1994 Apr (Peters - solid triangles), 1994 Jun -- 1997 Jun
(Richards, Albright \& Koubsk{\'y} - solid circles), 1999 Jan -- 2001 Jan
(Peters - solid triangles), and the 1956--1977 data from \citet{popper89}
(open circles).  The latest KPNO data from 1994--2001 (solid circles and
solid triangles) constitute the best data set.  The phases given here are
spectroscopic phases measured from periastron.  
} 
\label{f2} 
\end{figure*}

\begin{figure*}
\figurenum{3}
\epsscale{1.0}
\includegraphics[angle=-90,scale=0.6]{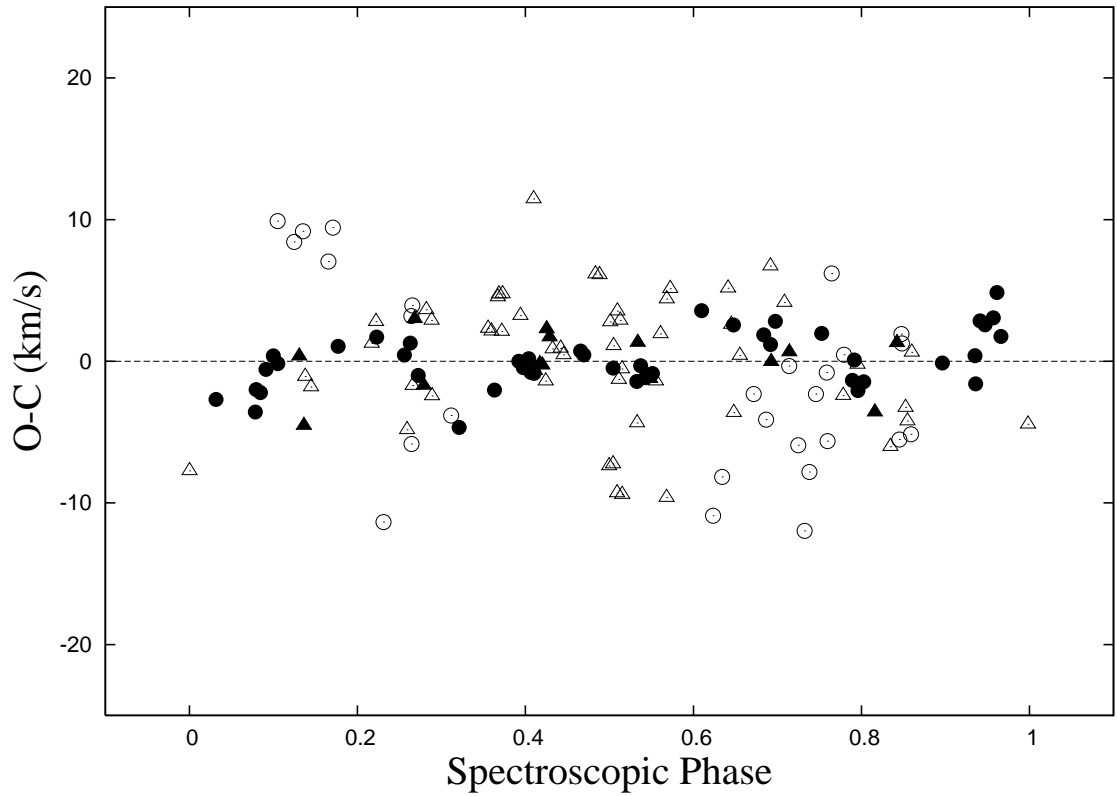}
\caption{
Residuals of the radial velocity curve of TT Hya (elements Table \ref{t4},
column 5), showing the KPNO spectra obtained from 1985 -- 2001 and the
data from \citet{popper89} (open circles).  The symbols have the same
notation as in Figure 2.  The phases given here are spectroscopic phases
measured from periastron. Note that the solid symbols representing the 
1994--2001 data have consistently lower O-C values than the other data.
}
\label{f3}
\end{figure*}

The differences in the orbital elements listed in Table \ref{t4} are very
small.  The orbital solution derived from our most accurate data
(1994--2001 data; column 2), is to be preferred.  However, the results
listed in column 5 provide the best orbit based entirely on spectroscopic
data.  We note that the orbital period of 6.953429 days derived by 
\citet{kulkarni+abhyankar80} is well outside the error bars of the new 
spectroscopic solution of $P = 6.953484 \pm 0.000012$ days.  
This spectroscopic orbital solution and radial velocity measurements are
illustrated in Figure \ref{f2}, and the (O-C) residuals are shown in
Figure \ref{f3}.  Figure \ref{f3} shows that Popper's measurements
\citep{popper89} tend to smooth the amplitude of the radial velocity curve
at the quadratures.

Assuming $a_{2}\sin i =1.2978\times 10^{7}$ km from the solution in Table
\ref{t4} column 2 based on the high-accuracy 1994--2001 data and an
inclination and mass ratio from \citet{vanhamme+wilson93}, we obtain
$a=23.04 R_{\odot}$ and $M_{1}+M_{2}=3.40 M_{\odot}$.  The stellar masses 
were computed directly from the derived semi-major axis using Kepler's Third Law, 
assuming a known mass ratio.  However, since \citet{vanhamme+wilson93} found that 
$a=22.63 R_{\odot}$,  we calculated improved stellar radii by rescaling the 
\citet{vanhamme+wilson93} radii by a factor of 23.04/22.63 or 1.018. This 
simple scaling of the system dimensions is necessitated by the absence of 
adequate new photometry, especially 
at red or near-infrared wavelengths.  Moreover, we hope that new simultaneous 
light curve and velocity curve solutions will be completed to further improve the 
parameters of this system.  The new radii and other derived system parameters are 
listed in Table \ref{t5}.     Here, the first column labeled ``VW'' lists the
results from Table 4b of \citet{vanhamme+wilson93}, while the second
column gives our improved values.  If we assume that $R/d=9.78$
\citep{plavec88}, where R is the radius of the primary in $R_{\odot}$ and
d is the distance to the system in kpc, then we obtain a new distance
$d=203$pc to TT Hya.  This new value is still not in good agreement with
the Hipparcos distance of 154 pc.  This means that either the Hipparchos
parallax was overestimated or the temperature of 9900 K assumed by
\citet{plavec88} was overestimated, i.e., his value of $R/d$ was
underestimated.

Table \ref{t4} also shows that the eccentricity differs from zero by more
than 7$\sigma$ and the standard deviation of the best photometric fit is
2.0 km\,s$^{-1}$, a significant improvement over the previous eccentricity
and circular orbit derived by \citet{popper89}.   An eccentric orbit for 
the secondary star could imply that the mass transfer process may not be 
uniform.  Based on our analysis of the spectra, this 
process could be intensified at periastron, when the Roche lobe shrinks, and 
attenuated at apastron.  Assuming that more material is released at periastron 
(photometric phase, $\phi = 0.537$) than at apastron, and that it takes time 
for this material to interact with the disk, get heated, and become luminous, 
then enhanced H$\alpha$ emission should be detected with the stronger blue peak 
at the phases shortly after $\phi = 0.55$.  Evidence of such enhanced emission 
was found at phases $\phi = 0.60-0.65$ (see \citealt{budaj+richards+miller05}, 
Figure 6); consequently enhanced emission near periastron is a viable possibility.   
Similar observations in other Algol-type binaries with non-zero eccentricities 
would provide stronger evidence of the association with periastron.  Moreover, 
these results may explain the RZ Cas system in which intensified pulsations of 
the primary star are strongest shortly after primary eclipse 
\citep{lehmann+mkrtichian04}, which occurs near periastron.

\subsection{Radial Velocity of the H$\alpha$ Central Absorption}

The H$\alpha$ line has a complicated shape because it contains double
peaked emission with a central absorption.  Table \ref{t2} lists our
measurements of the normalized fluxes of the blue and red emission peaks
($I(em_{b})$, $I(em_{r})$) and the central absorption
($I(ab_{c})$), as well as the radial velocity measurements of the
central absorption ($V_r(ab_{c})$).  The central absorption (or
depression) in the H$\alpha$ line profile is the strongest feature in our
spectra that can be ascribed (at least partly) to the primary star.

\begin{figure*}
\figurenum{4}
\epsscale{1.0}
\includegraphics[angle=-90,scale=0.6]{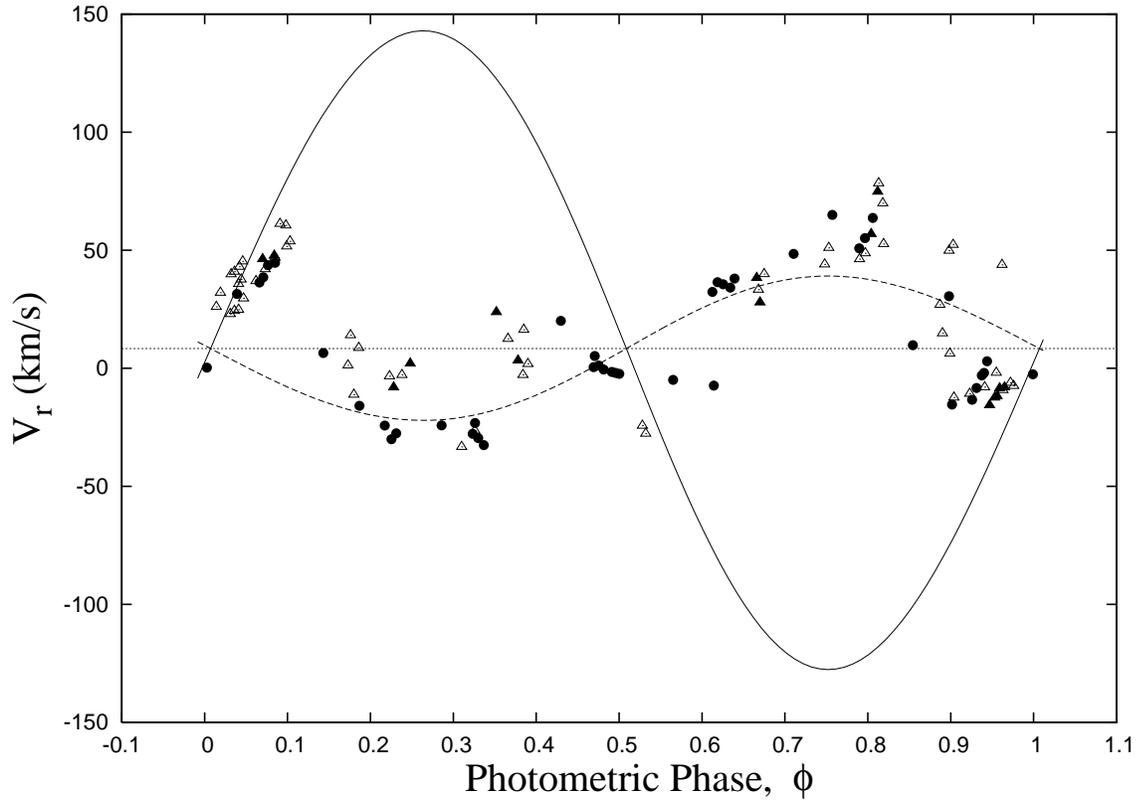}
\caption{
Radial velocity, $V_r(ab_{c})$, of the central absorption of the
H$\alpha$ line.  The solid line is the velocity curve of the secondary and
the dashed line is the velocity curve of the primary derived using the
\citet{vanhamme+wilson93} solution (see \S3.2).  The symbols have
the same notation as in Figure 2.  The phases shown are photometric
phases.
}
\label{f4}
\end{figure*}

The radial velocity of the H$\alpha$ line core, $V_r(ab_{c})$, is shown
in Figure \ref{f4}.  These measurements reflect the motion of the primary
star but are distorted by several effects at different phases.  Near first
quadrature (phase, $\phi \sim 0.1-0.4$), the spectrum shows blueshifted
absorption.  Closer to secondary eclipse ($\phi \sim 0.5$), we do not
observe any significant negative velocities and thus cannot confirm
previous suggestions of mass outflow at these phases \citep{peters89}.
The opposite quadrature from $\phi \sim 0.6-0.9$ has somewhat higher
redshifted velocities since the gas stream and disk which project onto the
primary star surface have a considerable radial, in-falling component of
the velocity.

Near primary eclipse the velocities of the central H$\alpha$ absorption
differ greatly from those predicted by the orbital motion of the primary.
It might be expected that the gradual occultation of the rapidly rotating
primary would considerably affect the measured radial velocities; however,
this rotational effect can only operate within the partial eclipse from
$\phi = 0.95-0.05$ and the velocity curve is distorted over a much
broader interval from $\phi = 0.9-0.1$.  Moreover, the rotational effect
would predict positive velocities from $\phi = 0.95-1.0$ and negative
velocities from $\phi = 0.0-0.05$, while the opposite trend is observed.
Another explanation of this effect is thus necessary.  The eclipse of the
approaching edge of the disk from $\phi = 0.9-1.0$ removes the
blueshifted emission peak, which should manifest as blueshifted absorption
in the observed H$\alpha$ profile (with an opposite effect for $\phi =
0.0-0.1$).  This phenomenon influences the broad shape of the H$\alpha$
core but is not the predominant cause of the anomalous radial velocity
measurements near primary eclipse.  It is the secondary star which is
mainly responsible for the effect.  Although its intrinsic H$\alpha$ line
is rather weak, it is narrow and can considerably affect the radial
velocity measurements of the center of H$\alpha$ absorption, particularly
at the phases when both components of H$\alpha$ (from the primary and
secondary) are aligned and become unresolved.  At these phases, the
secondary component of H$\alpha$ determines the deepest point of the whole
H$\alpha$ feature which was used in radial velocity measurements.  The
measurements of H$\alpha$ absorption near primary eclipse are in excellent
agreement with the velocity curve of the secondary, and similar
measurements performed on the synthetic spectra of the stars plus disk
also reveal a transition near primary eclipse where the H$\alpha$
absorption ceases to follow the velocity curve of the primary and instead
matches that of the secondary.  A qualitatively similar effect is observed
near secondary eclipse, although in this case the amplitude of the
measured radial velocities is slightly less than predicted by the
secondary velocity curve, presumably a result of the partial obscuration
of the secondary.

Future radial velocity measurements of the primary from weaker lines in
the infrared as well as R-band and infrared photometry (see e.g.,
\citealt{richards90}) might improve the parameters of the system
considerably, specifically the mass ratio and the temperature of the
secondary star.

\subsection{The ``REBECA'' Diagram}

\begin{figure*}
\figurenum{5}
\epsscale{1.0}
\plotone{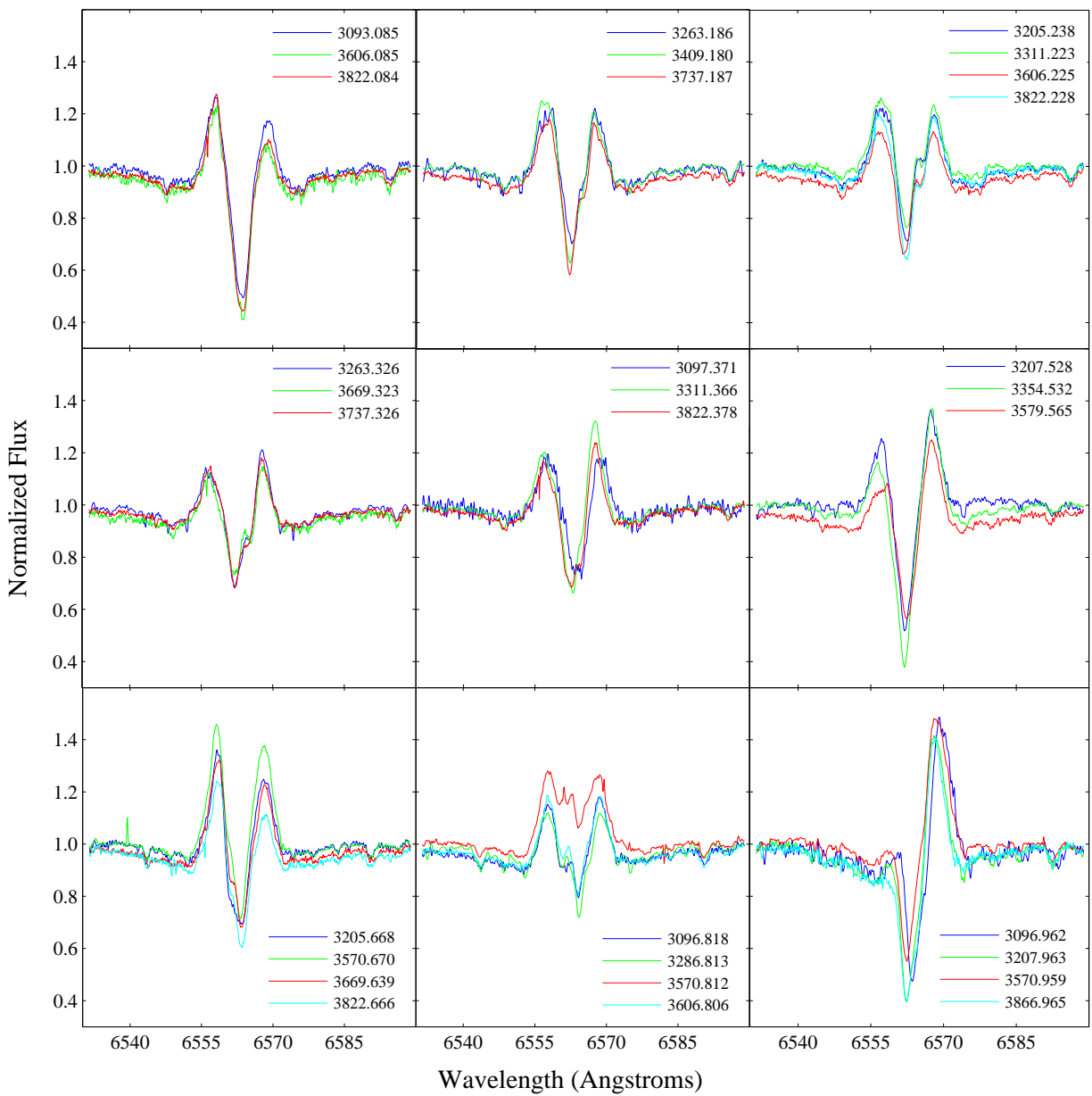}
\caption{
Comparison of the observed H$\alpha$ spectra at different epochs for
phases near $\phi$ = 0.08, 0.19, 0.23, 0.33, 0.37, 0.54, 0.67, 0.81, and
0.96.  There is remarkable similarity between the spectra at the epochs
separated by 833 cycles, except that the spectra at epoch 3570 had
significantly stronger disk emission than the other spectra.
}
\label{f5}
\end{figure*}

The variability of the H$\alpha$ line profile with epoch is illustrated in
Figure \ref{f5} for phases near $\phi$ = 0.08, 0.19, 0.23, 0.33, 0.37,
0.54, 0.67, 0.81, and 0.96.  This figure shows that the spectra obtained
in 1994 Apr (epoch 3570) had significantly stronger disk emission than the
other spectra.  The emission strength returned to normal by 1994 Jun 
(epoch 3579).

\begin{figure*}
\figurenum{6}
\epsscale{1.0}
\includegraphics[angle=-90,scale=0.6]{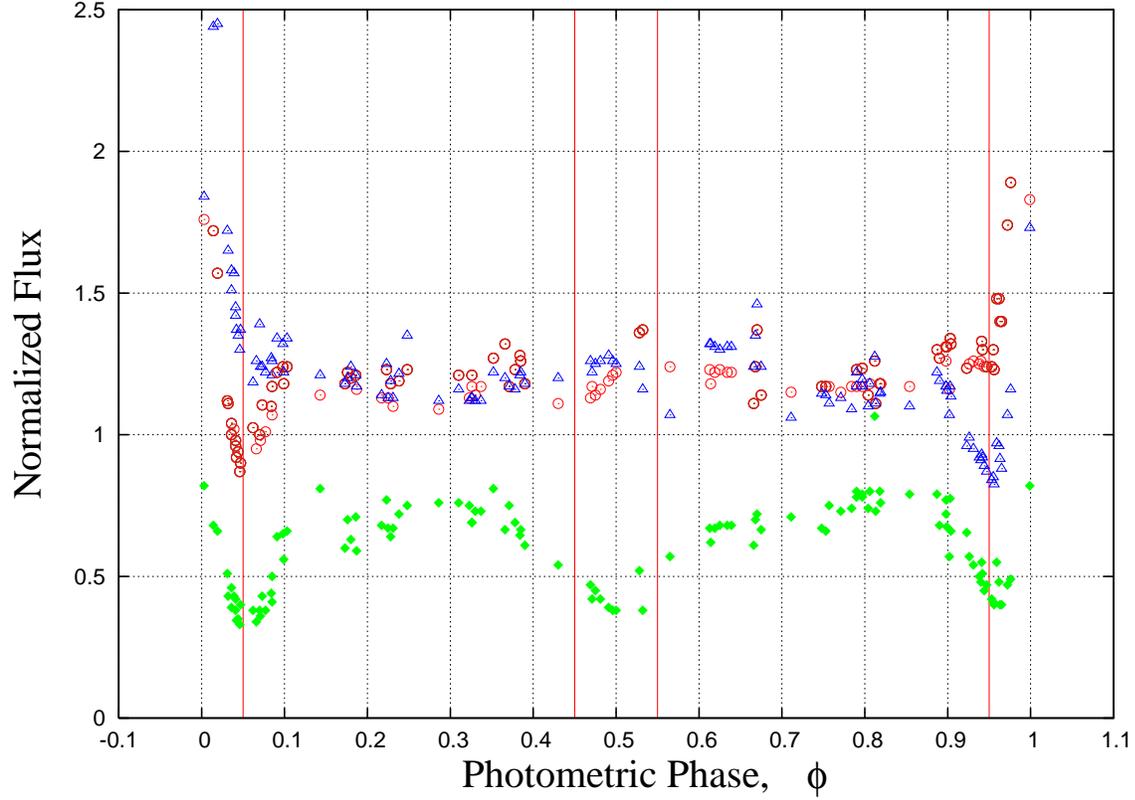}
\caption{
REBECA diagram for the 1985-2001 data showing the behavior of the emission
strength and central absorption depth around the orbit of TT Hya.  Blue
triangles -- highest normalized flux of the blue emission peak; red
circles -- flux of the red emission; and green diamonds -- depth of the
central depression.  The vertical lines indicate the start and the end of
primary and secondary eclipse.
}
\label{f6}
\end{figure*}

The three major characteristics of the complicated H$\alpha$ profile:
Red Emission, Blue Emission, Central Absorption ($I(em_{r})$,
$I(em_{b})$, $I(ab_{c})$) can be plotted as a function of phase,
$\phi$, in what we call a ``REBECA'' diagram.
\citet{budaj+richards+miller05} [BRM05] illustrated how various
structures could be identified in this diagram (see Figure 6 of BRM05).
We have revised their figure to include all TT Hya spectra from
1985--2001 (see Fig.  \ref{f6}).  Phases $\phi = 0.1-0.3$ show comparable
intensity in the blue-red emission peaks, while the red emission is
slightly stronger at $\phi = 0.3-0.4$.  BRM05 speculated that such an
effect, if accompanied by the reverse (blue emission stronger) during the
mirror phases ($\phi = 0.6-0.7$), could indicate the presence of
circumstellar matter in the vicinity of the secondary star between the
C1-C2 surfaces.  The new 1985--2001 observations at $\phi = 0.3-0.4$
confirm that the red emissions is stronger while at $\phi = 0.6-0.7$ the
blue emission is stronger.  However, the above mentioned idea of BRM05
probably cannot explain the effect since it would require a circumstellar
matter between C1-C2 surfaces projecting onto the surface of the
secondary with a considerable outflowing velocity component (relative to
the secondary star) at $\phi = 0.3-0.4$ and inflowing velocity component
during $\phi = 0.6-0.7$ which seems unlikely.  At $\phi = 0.4-0.5$, the
blue emission is stronger than the red emission and both emissions
increase slightly.  According to BRM05, this is caused by the extra
absorption of the secondary's light in the receding part of the disk and
by the eclipse of the secondary star.  The calculations of BRM05
predicted that the reverse should be seen during the mirror phases ($\phi
= 0.5-0.6$).  Our new observations from 1985 --2001 strongly confirm the
presence of this effect.  The central absorption gets progressively deeper
when approaching the secondary eclipse as described in BRM05.

The blue and red emission seen during $\phi = 0.7-0.88$ seem fairly
symmetric.  Starting at about $\phi = 0.88$, all three parameters
({$I(em_{b})$, $I(em_{r})$), ($I(ab_{c})$}) change dramatically.
The blue emission decreases rapidly due to the eclipse of the disk while
the red emission gets slightly stronger.  At about $\phi = 0.95$, both
emissions soar as we go into total eclipse and the reverse effects are
observed during $\phi = 0.0-0.1$.  This behavior was explained by BRM05,
however, the blue emission at about phase 0.02 seems surprisingly strong.
The ``W'' shape of the central absorption pattern near primary eclipse, 
from $\phi = 0.88-0.1$, was qualitatively explained by the calculations of 
BRM05.   In addition, they suggested the presence of cooler circumstellar 
matter around the secondary star which would produce an excess absorption 
in the core of H$\alpha$ at these phases. Using OI $\lambda$7774 spectroscopy,  
\citet{etzeletal95} identified similar interesting effects of disks and 
streams in several Algols around or near both eclipses.  We anticipate that 
synthetic spectra generated from the output of hydrodynamic simulations 
may more clearly explain the ``W'' behavior of the central absorption at 
these eclipse phases.

\subsection{Observed and Synthetic H$\alpha$ Spectra}

\begin{figure*}
\figurenum{7}
\epsscale{1.0}
\plotone{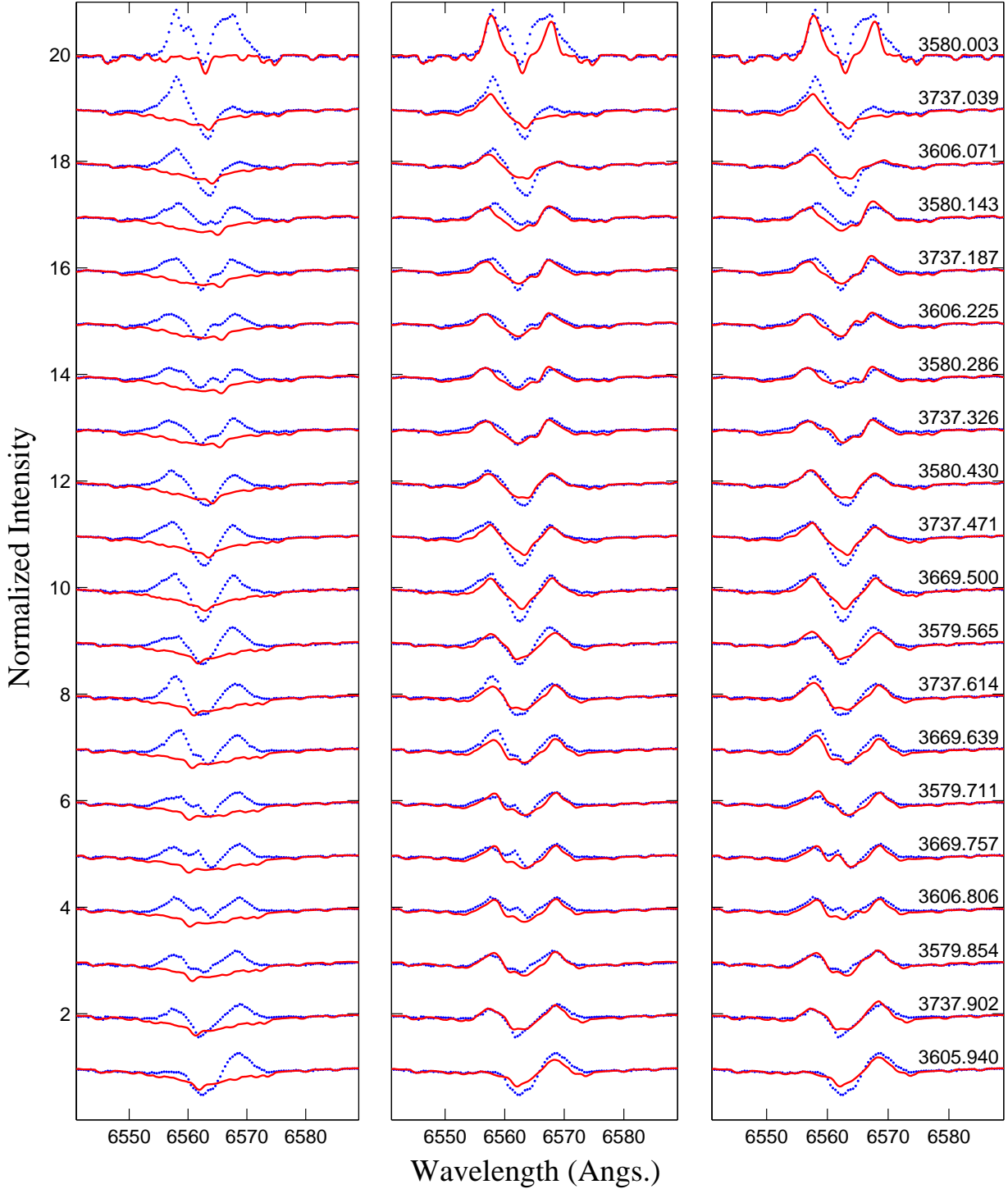}
\caption{
Comparison of the observed (dotted line) and synthetic (solid line)
H$\alpha$ spectra of TT Hya.  The epochs and orbital phases are listed to
the right of the figure.  The synthetic spectra include calculations of
(a) stars only (left frame); (b) stars and disk (middle frame); and (c)
stars, disk, and gas stream (right frame).  There is a significant
improvement in the fit between the observed and synthetic spectra once the
accretion disk is included, but only a minor improvement once the gas
stream is included in the model.
}
\label{f7}
\end{figure*}

The behavior of the line shape was studied in detail by BRM05.  In that
paper, we showed that the emission peaks originated from the disk and that
both the emission and central absorption varied with phase.  In addition,
we proposed a quantitative empirical model of the disk which could account
for the behavior of the observed H$\alpha$ line profiles.  However, BRM05
found that there remained a difference between the observed and synthetic
spectra which could be improved if the primary star was assigned a lower
effective gravity than derived earlier and the gas stream was considered
in the calculations.  In this paper, we have recalculated the synthetic
spectra by including the gas stream as well as the stars and accretion
disk in the model.  These changes improved the fits between the observed
and synthetic spectra at all phases outside of primary eclipse, $\phi = 0.225
- 0.854$ (see Fig. \ref{f7}), while having almost no effect on the
primary eclipse spectra.

The {\sc{synspec}} code \citep{hubenyetal94} was used to calculate the new
intrinsic (not rotationally broadened) spectrum of the primary assuming a
lower surface gravity of $\log g=3.5$, microturbulence of $v_{\rm trb}=2$
km\,s$^{-1}$, solar abundances, as well as Kurucz models and line lists
\citep{kurucz93a, kurucz93b}.  The synthetic spectrum of the secondary was
based on the parameters used in \citet{budaj+richards+miller05}.  New
composite synthetic spectra of the stars, disk, and gas stream, were
calculated using the {\sc{shellspec}} code \citep{budaj+richards04}, for
phases corresponding to the observations.  The gas stream was treated as a
simple cylinder located along the predicted freefall path with
approximately freefall velocities.  Since the disk extends almost to the
Roche lobe and has a thickness of $2R_{\odot}$ above (and below) the
orbital plane, the radius of the stream cylinder was set to $R=3R_{\odot}$
to produce an overlap between the gas stream and disk locations.  This
overlap permitted the stream to flow through only above and below the
disk.  The density, temperature and velocity field of the disk were
assigned a higher priority than the gas stream in the overlap region.

Initially, a gas stream temperature of $T = 8000$K was chosen to
correspond to the peak in H$\alpha$ emissivity.  The estimated density of
the stream at the starting point ($2\times10^{-14}$g\,cm$^{-3}$) was
chosen to improve the fit between the observed and synthetic emission at
the quadratures of the orbit and during the eclipses.  For these
calculations, the density was allowed to vary along the stream to satisfy
the continuity equation.  The density of the disk at the inner radius was
$3.3\times10^{-14}$g\,cm$^{-3}$ and the temperature of the disk was
$T=7000$K.  Note that this density derived from the H$\alpha$ line is very 
close to the value of $5\times10^{-14}$g\,cm$^{-3}$ obtained from the IUE spectra 
(see \S4 and Table \ref{t6}), but the IUE solution produced a better fit to the data.  
The electron number densities in the disk and stream were
calculated from the temperature, density, and chemical composition using
the Newton-Raphson linearization method.  All the other parameters of the
system were the same as in Table 1 of \citet{budaj+richards+miller05}.
The comparison between the observed and synthetic spectra in Figure
\ref{f7} demonstrates that the inclusion of the gas stream in the
synthetic spectra slightly improves the fit between observation and
theory, particularly near the quadratures.

The mass transfer rate can be estimated from the velocity, density, and
cross-section (the two semicircular sectors) of the gas stream above and
below the disk.  We obtained a rate of
$\sim2\times10^{-10}M_{\odot}yr^{-1}$, two orders of magnitude higher than
the lower limit of $10^{-12}M_{\odot}yr^{-1}$ obtained by
\citet{peters+polidan98}.  Since the temperature of the stream was chosen
to correspond to the peak H$\alpha$ emissivity, our value of the mass
transfer rate is still a lower limit.  This rate depends essentially on
the equivalent width of the excess emission corresponding to the stream,
i.e., on the stream temperature and density, but not significantly on any
particular choice of stream geometry.

\section{The Ultraviolet Spectra}

The ultraviolet spectra of TT Hya analyzed in this work were derived from
IUE spectra collected from 1980 - 1992 (see \S 2).  A few of these
spectra were studied by \citet{plavec88}, but his analysis focussed on the
low resolution spectra and included only two out-of-eclipse high
resolution IUE spectra.  Since that time, seventeen high resolution
spectra were obtained with the IUE telescope, including two during total
eclipse.  These data are summarized in Table \ref{t3}.  Only the high
resolution spectra have been analyzed in this work since the low
resolution spectra are not suitable for our radial velocity studies.

\subsection{Radial Velocities of the Primary from UV lines}
\label{s4}
In the optical spectra, there are very few absorption lines of the primary
and these are often heavily distorted and contaminated by emission
features.  However, the ultraviolet lines can be used to measure radial
velocities, which are crucial for establishing a precise mass ratio, and
thus the geometry of the system (see e.g., \citealt{baraietal04}).  For
this reason, radial velocities of the primary were measured from the high
resolution IUE spectra.

\begin{figure*}
\figurenum{8}
\epsscale{1.0}
\plotone{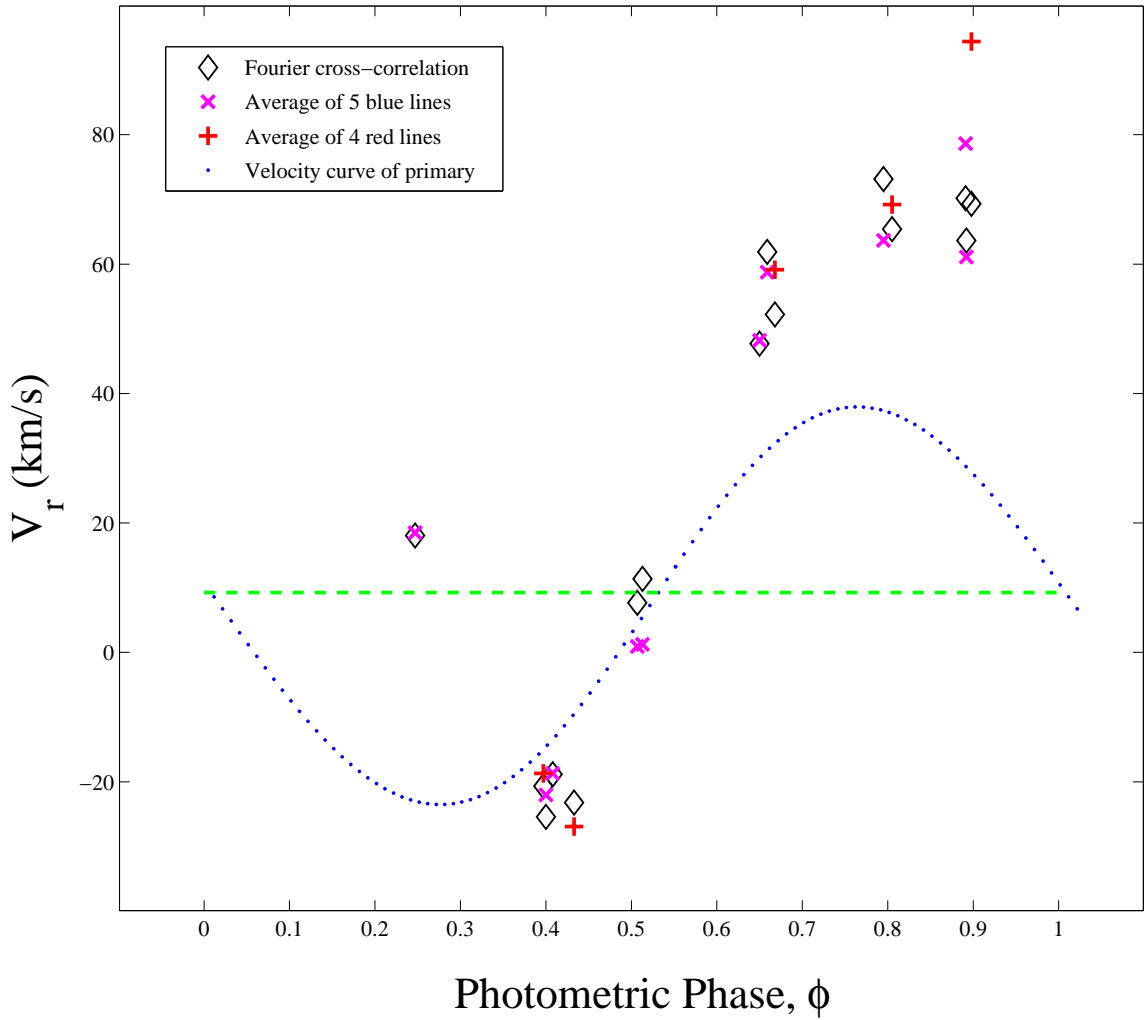}
\caption{
Radial velocities of UV lines compared with the velocity curve of the
primary star (dotted line).  The deviation from this curve suggests that
the measured UV lines do not originate from the primary but from a
loosely-bound structure with non-circular velocities, like an asymmetric
or elliptical disk around the primary star.
}
\label{f8}
\end{figure*}

First, we calculated the synthetic spectrum of the primary with the
{\sc{synspec}} code \citep{hubenyetal94}.  Once again, Kurucz models and
line list, with vacuum wavelengths, were used \citep{kurucz93a,
kurucz93b}.  Since the observed ultraviolet lines were narrower than
expected from the rotational velocity ($v\sin i = 168$ km\,s$^{-1}$; Table
\ref{t6}), the synthetic spectrum was convolved with a rotation profile of
80 km\,s$^{-1}$, to improve the accuracy of the cross correlation
procedure.  (Note that the slightly lower rotational velocity is
consistent with our later suggestion (\S4.2) of a circumstellar origin for
the UV lines.)  The radial velocities were then obtained via
cross-correlation of the observed and synthetic spectra.  These velocities
are listed in Table \ref{t3} and displayed in Figure \ref{f8}.
Qualitatively, except at phase 0.25, the ultraviolet spectrum moves in
phase with the primary star.  However, quantitatively this trend does not
overlap with the expected motion of the primary.  \citet{plavec88}
suggested that the UV lines did not originate from the primary star but
from the accretion disk.  We confirm that they arise from a loosely-bound,
cooler structure with non-circular velocities, like an elliptical
accretion disk around the primary star.

Figure \ref{f8} can also be used to speculate on the geometry and dynamics
of the disk.  Starting at $\phi = 0.9$, we observe the highest radial
velocities corresponding to the highest infall velocities of about
40km\,s$^{-1}$, which suggests that the ``stream''-lines in the disk are
not fully circular but are influenced by the infalling gas stream.  This
infalling velocity component decreases as we move back in phase and
disappears somewhere in the region of $\phi = 0.5-0.6$.  This indicates
that the streamlines are almost tangential at these phases and that the
disk shape may be elliptical, with periastron at $\phi = 0.5-0.6$.  At
earlier phases, $\phi \sim 0.4$, we observe velocities more negative than
expected which indicates outflowing streamlines in agreement with the
elliptical disk model.  The only measurement which does not concur with
this model is the one at $\phi = 0.25$, which was obtained during the same
epoch (3363; Table 2) as several other points in Figure \ref{f8}.

\subsection{Synthetic and Observed Iron Curtain Spectra Outside 
of the Eclipses}
\label{s5}

The ultraviolet spectrum of an Algol-type binary is dominated by the light
of the primary star (except during total primary eclipse) since the
contribution from the cooler secondary star is insignificant at these
wavelengths.  The UV region in TT Hya contains many absorption-dominated
spectral lines.  However, these lines do not originate from the primary
star but probably from an accretion disk as suggested by \citet{plavec88}
based on their narrow widths and unusual depths.

Most of the lines in the UV spectrum are blended Fe \ii\ lines which are
often referred to as the ``iron curtain'' \citep{shore92}.  The ``iron
curtain" origin of the UV lines is confirmed by their radial velocities
(see \S\ref{s4}) which are decoupled from the velocities of the
primary star.  These sharp and deep absorption lines were used to derive
an independent estimate of the temperature of the accretion disk.  We
modeled the region of the UV spectrum in the vicinity of the Al\iii\
$\lambda$1854 resonance doublet from 1850 \AA\ to 1867 \AA.  The following
nine spectral lines were used to derive the final model:
Fe\ii\ $\lambda$1854.659,
Al\iii\ $\lambda$1854.716,
Fe\ii\  $\lambda$1859.746,
Fe\ii\  $\lambda$1860.053,
Fe\ii\  $\lambda$1860.132,
Al\ii\  $\lambda$1862.478,
Al\iii\ $\lambda$1862.790,
Fe\ii\  $\lambda$1864.647, and
Fe\ii\  $\lambda$1864.749. 

As in the case of the optical spectra, the synthetic spectrum of the
primary star was calculated using the {\sc{synspec}} code
\citep{hubenyetal94}.  Solar abundances, Kurucz models and line list with
vacuum wavelengths \citep{kurucz93a, kurucz93b}, and a microturbulence of
2 km\,s$^{-1}$ were assumed.  This spectrum was then assigned to the star
and a complex spectrum of the star and disk in absolute units was
calculated with the {\sc{shellspec}} code \citep{budaj+richards04}.  The
geometry of the primary star and disk were taken from
\citet{budaj+richards+miller05}, and our derived distance of 203 pc was
assumed in the calculations.  The density of the disk was also adopted
from \citet{budaj+richards+miller05} since these values were obtained from
the H$\alpha$ emission strength, which provides a good measure of the
density.  The electron number density was derived using the Newton-Raphson
linearization and iteration method from the temperature, density, and
abundances.  The initial temperature of the primary star ($\teff=9900$K)
was previously derived from the UV continuum by \citet{plavec88}.
However, the temperature of the star and disk were varied to match the
observed IUE spectra.

The best fit to the local continuum was obtained from the Kurucz models
for a stellar temperature $\teff=10000$K, which is in very good agreement
with the result of \citet{plavec88}.  The Al\iii\ and Fe\ii\ lines in this
region are sensitive temperature indicators, and a disk temperature of
$T=7000$K provided the best fit to the UV lines.  At cooler temperatures,
the Al\iii\ lines in the synthetic spectra disappear, while hotter
temperatures lead to a disappearance of the Fe\ii\ lines in the synthetic
spectra.  This new disk temperature is in good agreement with the estimate
of $T=6200$K found by \citet{budaj+richards+miller05} from the H$\alpha$
line.  In addition, the new disk temperature is $\sim \frac{2}{3}\teff$ of
the primary star, which is a common assumption in the case in Be star
disks \citep{{hummel+vrancken00}, {vankerkwijk+waters+marlborough95}} and
which may indicate that the Be disk phenomenon may extend to the
Algol-type binaries.

Thomson scattering in the disk does not affect the spectrum significantly
since the disk is optically thin in the continuum.  In the center of
strong lines, like the Al\iii\ 1854.7 line, the opacity can be 3-4 orders
of magnitude higher than in the continuum and the optical depth can
sometimes be greater than 10.  Under these circumstances, the assumption
of an optically thin medium in the treatment of scattering is not
fulfilled within the {\sc{shellspec}} code.  However, in LTE, the line
source function is the Planck function, thus the scattering does not
affect the lines at all.  Table \ref{t6} summarizes the parameters of the
star and disk derived from this UV region. 

\begin{figure*}
\figurenum{9}
\epsscale{1.0}
\epsfig{file=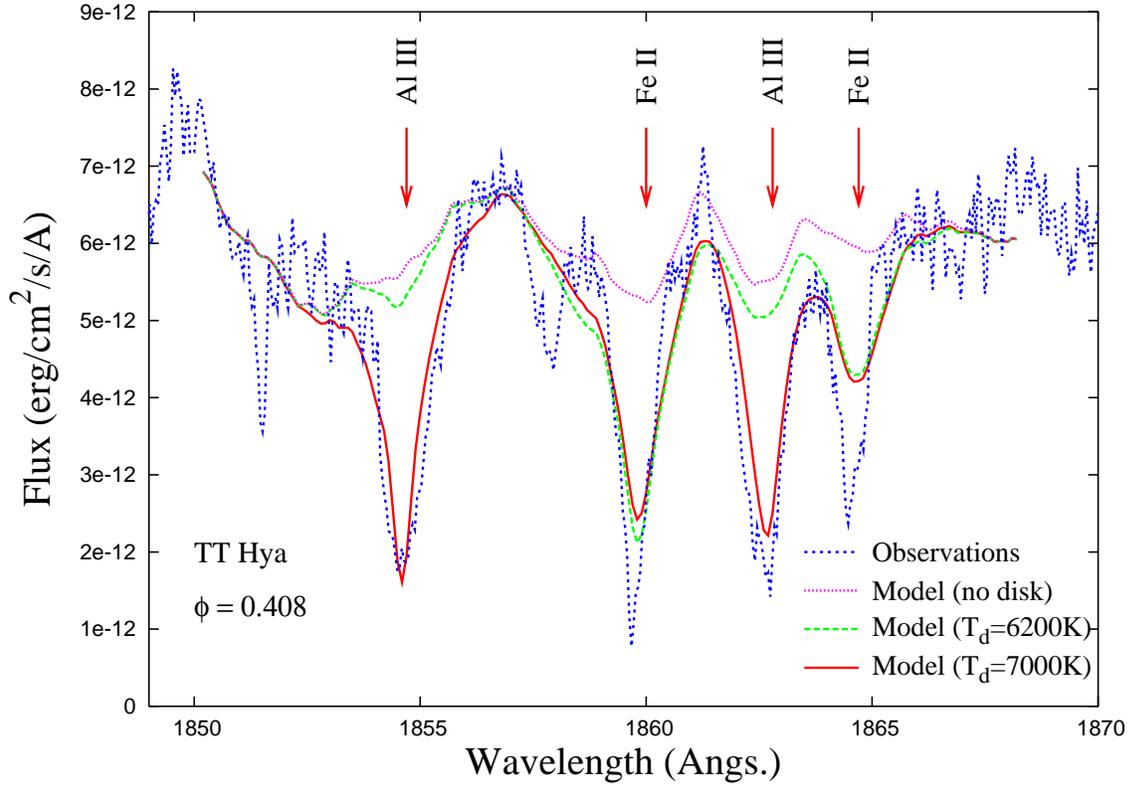,width=11cm,angle=-90}
\caption{
Observed and synthetic ultraviolet spectra in absolute flux units in the
vicinity of the Al\iii\ resonance lines.  Calculations demonstrate that
the UV absorption lines seen outside of the eclipses do not originate from
the primary star but mainly from the accretion disk, which has a
characteristic temperature, $T_d \sim 7000$K.
}
\label{f9}
\end{figure*}

Figure \ref{f9} clearly demonstrates that the synthetic stellar spectrum
of the primary cannot fit the UV absorption lines.  Once the disk is
included, a disk temperature of $T=7000$K provides a better fit to the
observed Al\iii\ lines than a temperature of $T=6200$K.

\subsection{Synthetic and Observed UV Spectra During Total Eclipse}
\label{s6}
\begin{figure*}
\figurenum{10}
\epsscale{1.0}
\epsfig{file=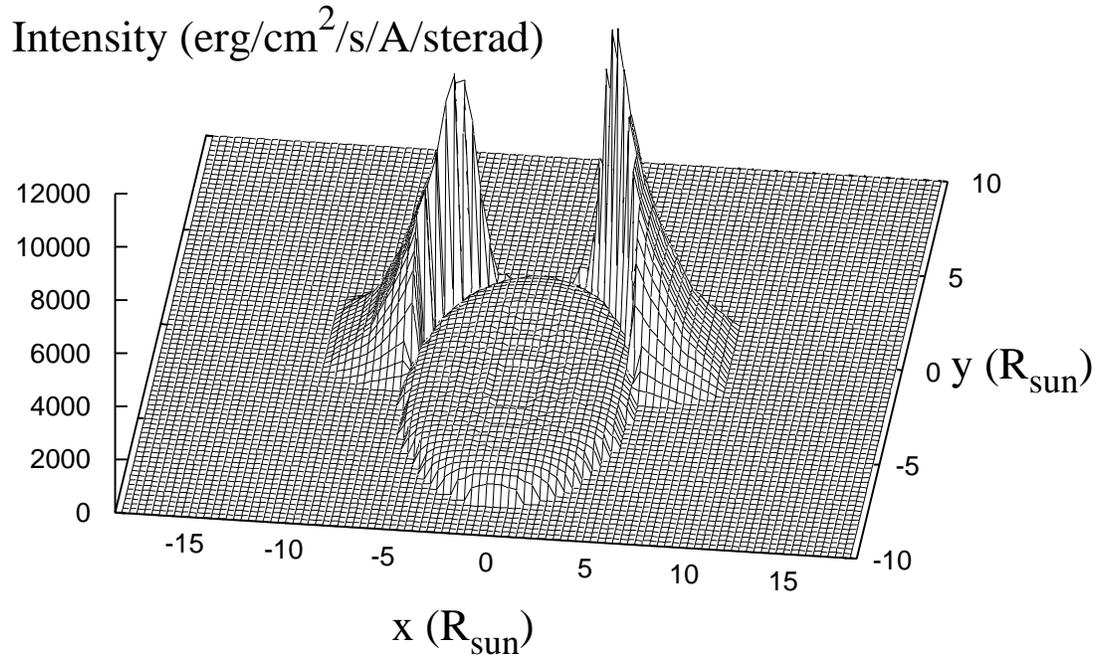,width=13cm,angle=-90}
\caption{
2D projection image of TT Hya in the continuum near 1850\AA\ during total
eclipse ($\phi = 0.001$).  The central rounded structure represents the
image of the secondary star.  Its rounded shape results from limb
darkening, while the depression in the center of the secondary star is due
to the gravity darkening.  The large peaks on either side of the secondary
star represent the disk emission, which is mainly due to Thomson
scattering of the light from the primary star.  The primary star is hidden
behind the image of the secondary.
}
\label{f10}
\end{figure*}

A total eclipse of the primary star provides a unique opportunity to study
the energy balance between the stars and circumstellar matter, the effects
of various radiative processes, and the geometry of the circumstellar
material.  For this reason, we used {\sc{shellspec}} to calculate the
total absolute flux in the continuum and lines during total eclipse (at
$\phi = 0.001$) in the spectral region containing the Al\iii\ lines (1850
\AA -- 1867 \AA).  The parameters of the disk and primary star were
assumed from the previous section.  However, the secondary star had to be
included in the calculation of the spectra seen during total primary
eclipse.  In the calculation of the synthetic spectra, the secondary was
considered to be a blackbody with a temperature of $T=4600$K at the
rotation poles.  Roche geometry, limb darkening, and gravity darkening
were included.  At $\lambda\approx 1850$\AA\, the model based on two stars
and a disk predicts a total absolute {\it continuum} flux of
$F_{\lambda}=3\times 10^{-15}$ erg\,cm$^{-2}$\,s$^{-1}$\AA$^{-1}$.
Approximately one half of this flux comes from scattered light in the disk
arising from the primary, while thermal radiation of the disk is
negligible (on the order of 1\%).  The continuum flux from the secondary
star is comparable to that of the scattered light.  Figure \ref{f10}
illustrates a 2D projection image of the system (as if seen on the sky) at
mid-primary eclipse ($\phi = 0.001$) for wavelengths near 1850~\AA~in the
continuum.  In this figure, the image of the secondary star is rounded
because of limb darkening, while the depression in the center of the
secondary star is due to gravity darkening.  The large peaks on either
side of the secondary star represent the disk emission, which is mainly
due to Thomson scattering.

\begin{figure*}
\figurenum{11}
\epsscale{1.0}
\epsfig{file=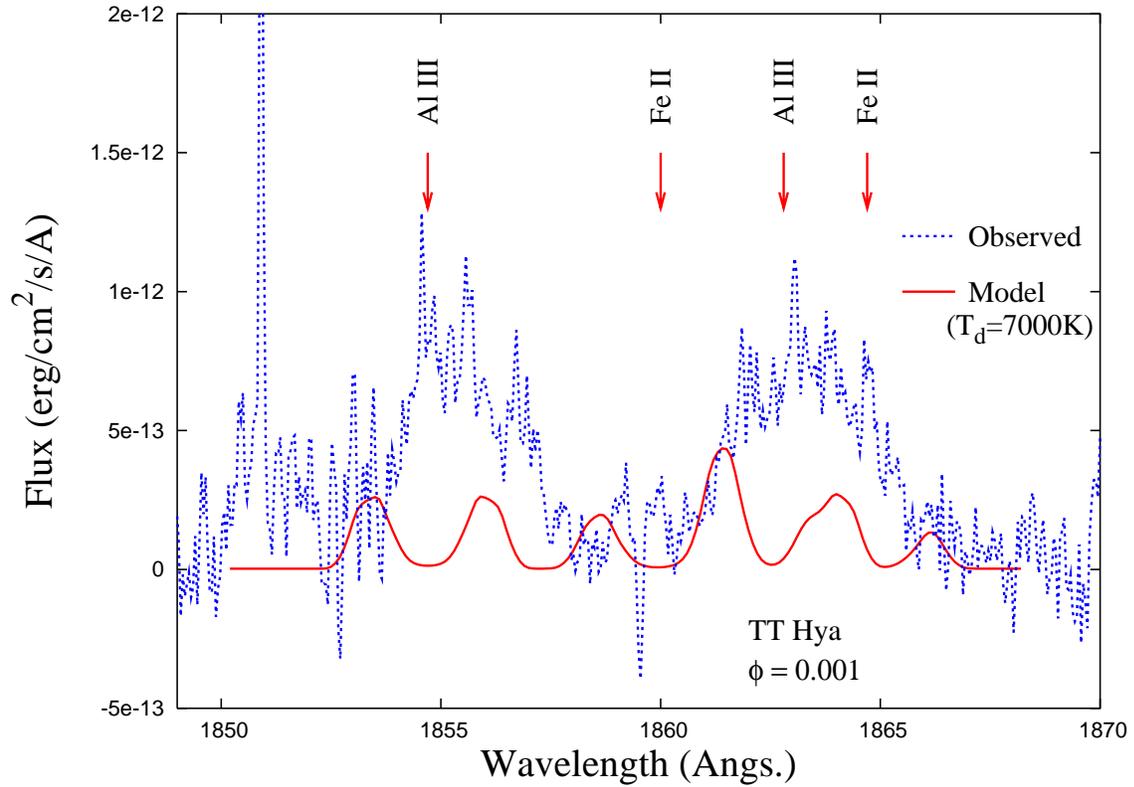,width=11cm,angle=-90}
\caption{
Short wavelength region of the IUE SWP high resolution spectra of TT Hya
during total eclipse (dotted line) compared with the synthetic spectrum
(solid line) which includes both stars and the disk (disk temperature,
$T_d = 7000K$).  It is clear that the disk cannot account for the emission
lines of higher ions.
} 
\label{f11}
\end{figure*}

The fit between the observed and synthetic spectra at mid-primary eclipse
is illustrated in Figure \ref{f11}.  The lines that were seen in
absorption outside of the eclipses become double-peaked emission lines
during total eclipse.  At lower resolution, these lines would become
unresolved and should create a pseudo-continuum with a flux of $\sim
F_{\lambda}=1\times 10^{-13}$erg\,cm$^{-2}$\,s$^{-1}$\AA$^{-1}$ level.
This is in agreement with \citet{plavec88} who determined the
pseudo-continuum flux of $1.2\times 10^{-13}$
erg\,cm$^{-2}$\,s$^{-1}$\AA$^{-1}$ from low resolution IUE spectra at this
wavelength.  However, it is clear from Figure \ref{f11} that a disk with a
temperature of $T=7000$K cannot account for the emission seen in lines of
higher ions like Al\iii\ during primary eclipse.  A hotter emission
source, with geometry and dynamics different from an accretion disk, is
required to explain the observed spectra.  The most important opacity
source in the continuum at this phase is Thomson scattering.  This
opacity source is typically two orders of magnitude higher than the
hydrogen bound-free opacity, which is a few times higher than the
Rayleigh scattering on neutral hydrogen; and the Rayleigh scattering is
about 2-3 orders of magnitude higher than the hydrogen free-free opacity.

\citet{lynchetal96} carried out spectropolarimetric observations of TT Hya
in the spectral region from 1600--2300\AA\ during total eclipse.  They
reported the presence of a non-stellar continuum which was considerably
polarized (20-30\% at the shortest wavelength).  Based on previous
arguments, this continuum is a pseudo-continuum created by emission in the
lines.  The fact that this light is polarized indicates that the line
emission is not fully thermal but that there is scattering in the lines as
well.  This finding suggests that the real source function in the lines is
not fully a Planck function and that there are departures from LTE, which
leads to non-LTE effects.

\subsection{The Si\iv, C\iv, Al\iii, and Mg\ii\ Emission Lines During 
Total Eclipse}
\label{s7}

\begin{figure*}
\figurenum{12}
\epsscale{1.0}
\epsfig{file=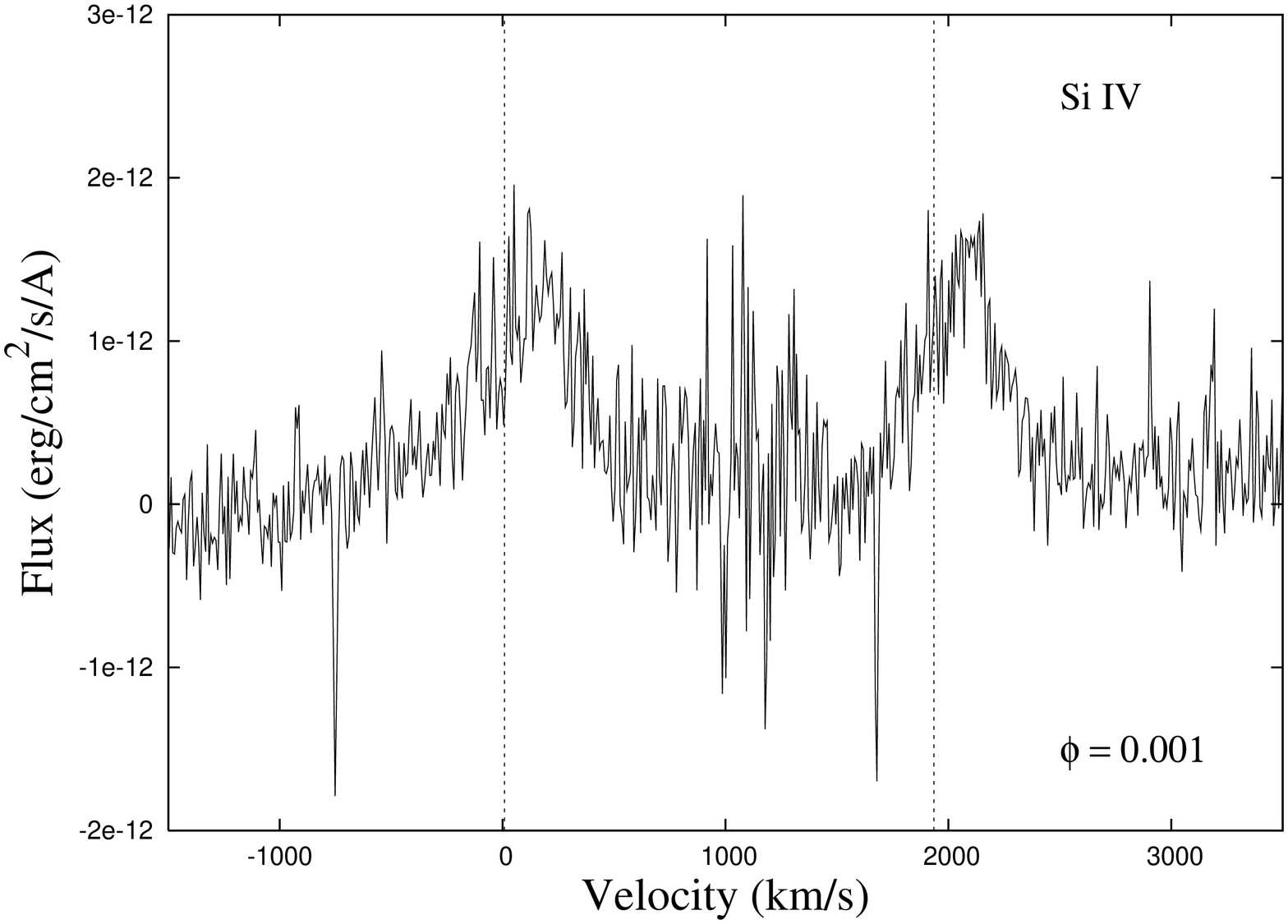,width=5.5cm,angle=-90}
\epsfig{file=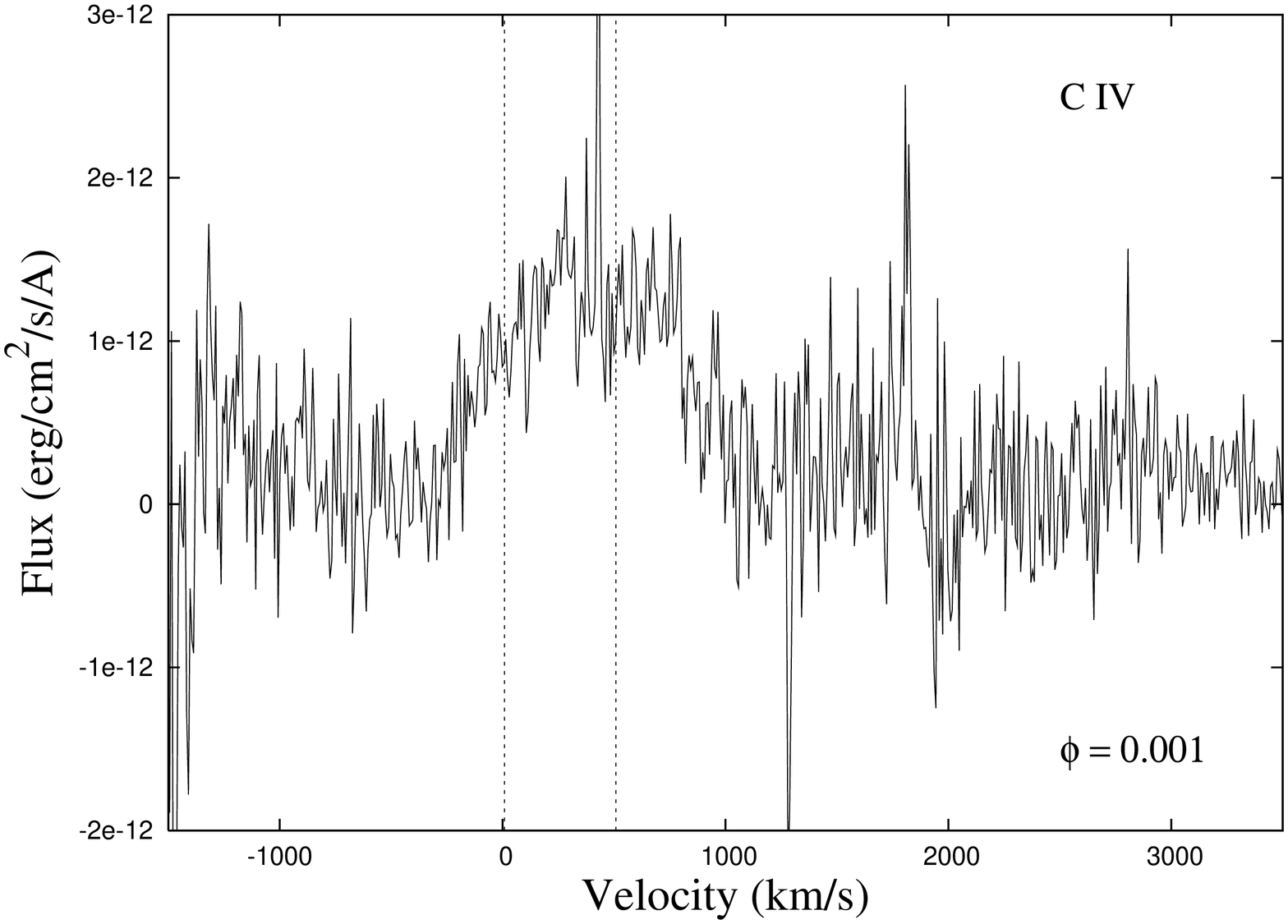,width=5.5cm,angle=-90}

\epsfig{file=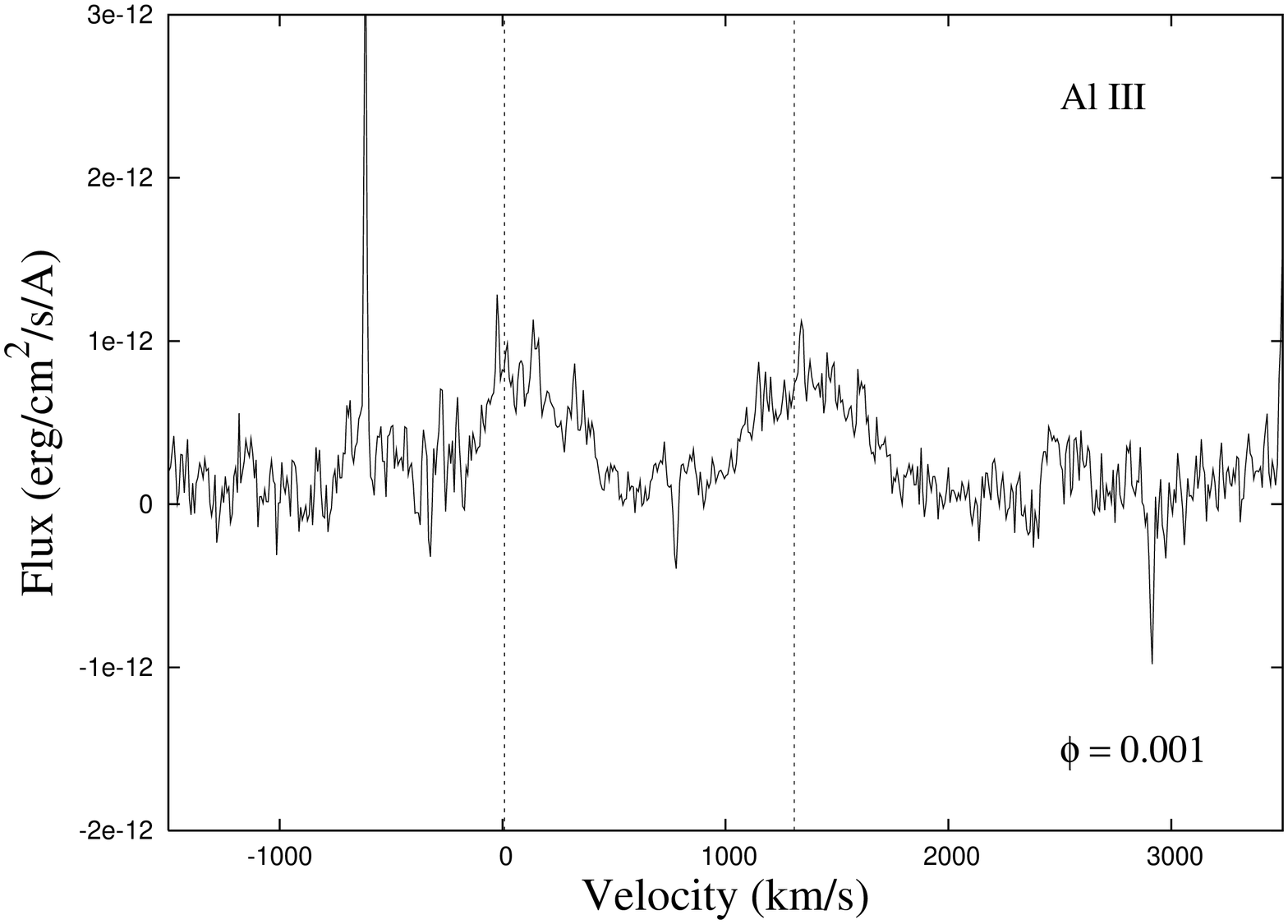,width=5.5cm,angle=-90}
\epsfig{file=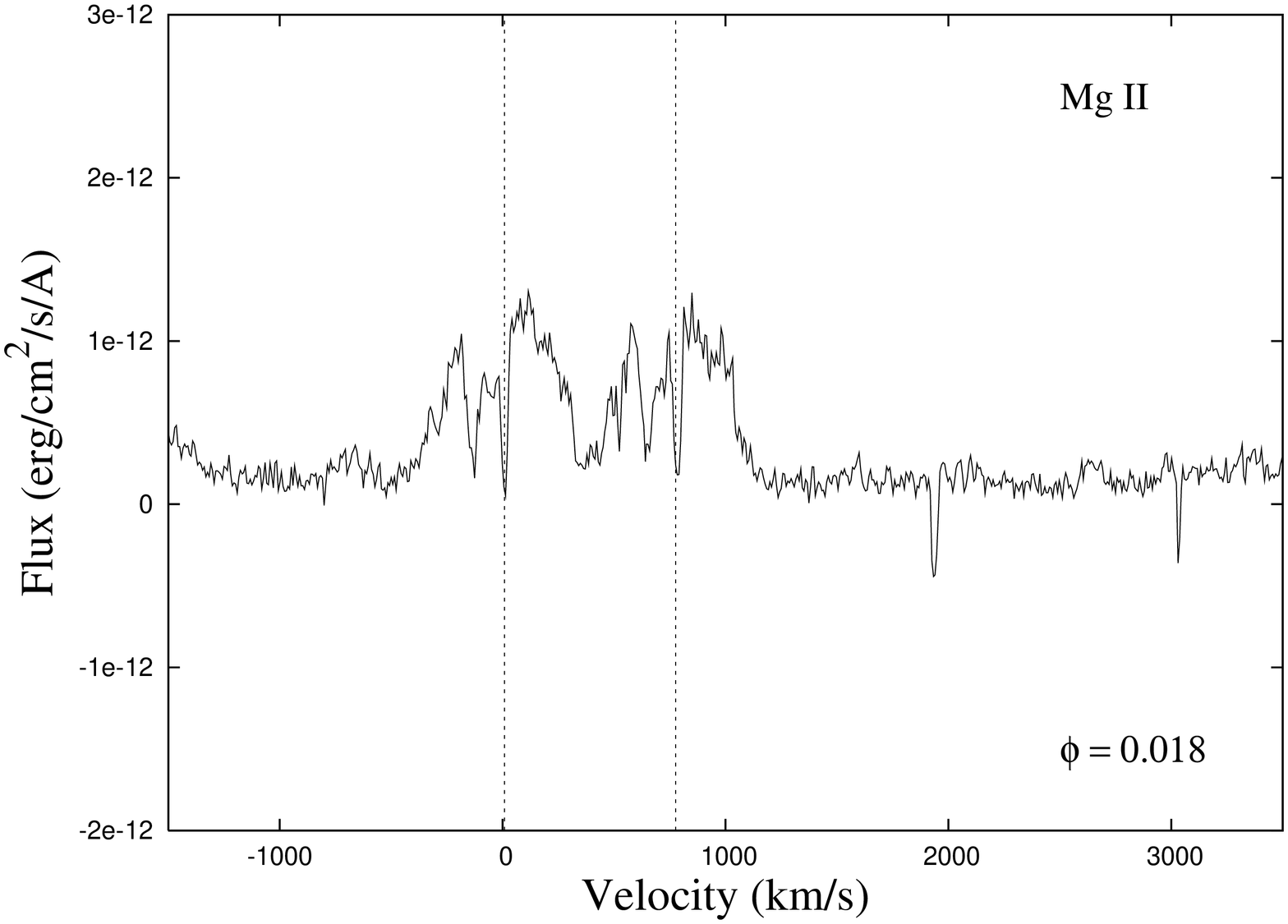,width=5.5cm,angle=-90}
\caption{
The Si\iv\ ($\lambda$1394, $\lambda$1403), C\iv\ ($\lambda$1548,
$\lambda$1551), Al\iii\ ($\lambda$1855, $\lambda$1863), and Mg\ii\
($\lambda$2796, $\lambda$2804) doublets during the total eclipse of the
primary in TT Hya.  The vertical lines correspond to the wavelength in a
vacuum of the doublet lines shifted to the mass center of the system.  The
Si\iv, C\iv, Al\iii\ lines were obtained at $\phi = 0.001$, and the Mg\ii\
line at $\phi = 0.018$.
}
\label{f12}
\end{figure*}

During the total eclipse of the primary star, various emission lines of
highly ionized species can be identified.  \citet{plavec88} could not
explain the source of this ionization.  Figure \ref{f12} illustrates the
high resolution IUE spectra of the Si\iv\ ($\lambda$1394, $\lambda$1403),
C\iv\ ($\lambda$1548, $\lambda$1551), Al\iii\ ($\lambda$1855,
$\lambda$1863), and Mg\ii\ ($\lambda$2796, $\lambda$2804) doublets during
primary eclipse.  Each emission line has a single-peaked profile, which
suggests that this emission does not originate in a disk.  The line
centers (except for Mg\ii) are redshifted by about 100-200 km\,s$^{-1}$
and the line widths extend to $\pm 300$ km\,s$^{-1}$ from the line
centers.

Comparison with our synthetic spectra also suggests that the disk
temperature ($T=7000$K) is too low to produce these lines.  Moreover,
these lines cannot originate on the primary or in the inner part of the
disk since these regions are eclipsed at this phase.  So, these
highly-ionized lines must originate from a region with considerable
vertical dimension above (or below) the orbital plane.
\citet{peters+polidan04} arrived at the same conclusions based on their
studies of high-to-moderate ionization emission lines in the far UV
obtained with FUSE.  They suggested the presence of a bipolar jet to
account for these features.  However, their interpretation may not provide
a sufficient explanation for the observed IUE spectra.  Symmetric bipolar
jets perpendicular to the orbital plane and with an orbital inclination of
$83^{\circ}$ would lead to jet velocities of $\sim$ 1200 km\,s$^{-1}$.
One jet would have to be occulted during the eclipse since we observe only
one redshifted feature.  Thus the jets would have to be shorter than about
7 $R_{\odot}$.  Moreover, we should expect an occulted jet to display
blueshifted emission based on geometric arguments, rather than the
redshifted emission that was observed.  The jets would also have to be
highly collimated, with opening angles $90 - i \sim 10$ degrees to avoid
smoothing the observed redshifted emission.  Consequently, there are many
constraints on the jet model.  We propose, instead, that the emission
arises from a hot disk-stream interaction region and an extended region
near the impact site.  This region would have a redshifted radial velocity
component of the correct size during total eclipse, and would naturally
have higher temperatures than the disk.  The structure would also extend
below and above the disk plane due to the disk-stream interaction and be
turbulent with supersonic velocities.  The region would be un-occulted
near third contact, which would explain the behavior of O{\sc vi} detected
by \citet{peters+polidan04}.

\subsection{The Ly$\alpha$ Line} 
\label{s8}

\begin{figure*}
\figurenum{13}
\epsscale{1.0}
\epsfig{file=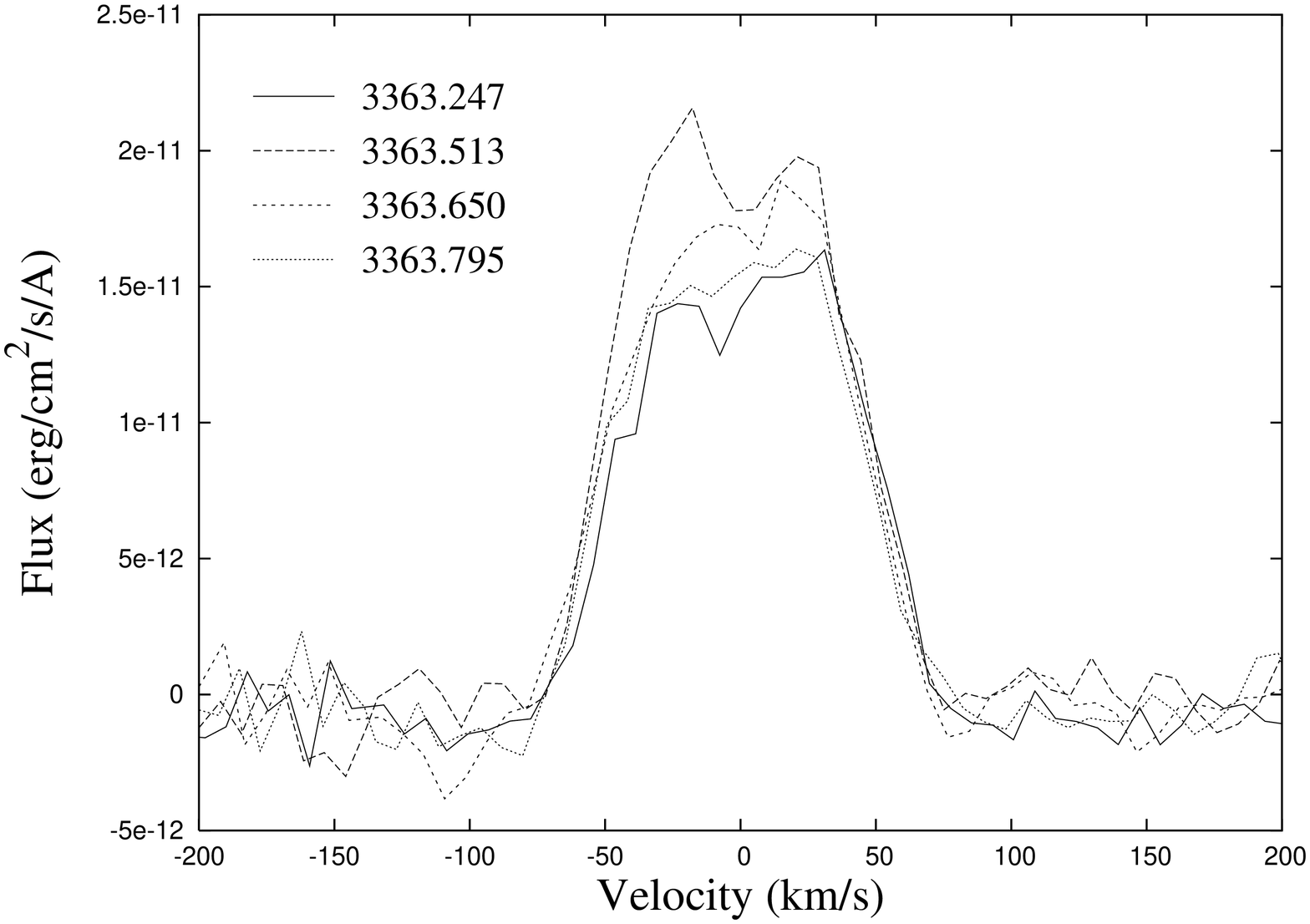,width=5.5cm,angle=-90}

\epsfig{file=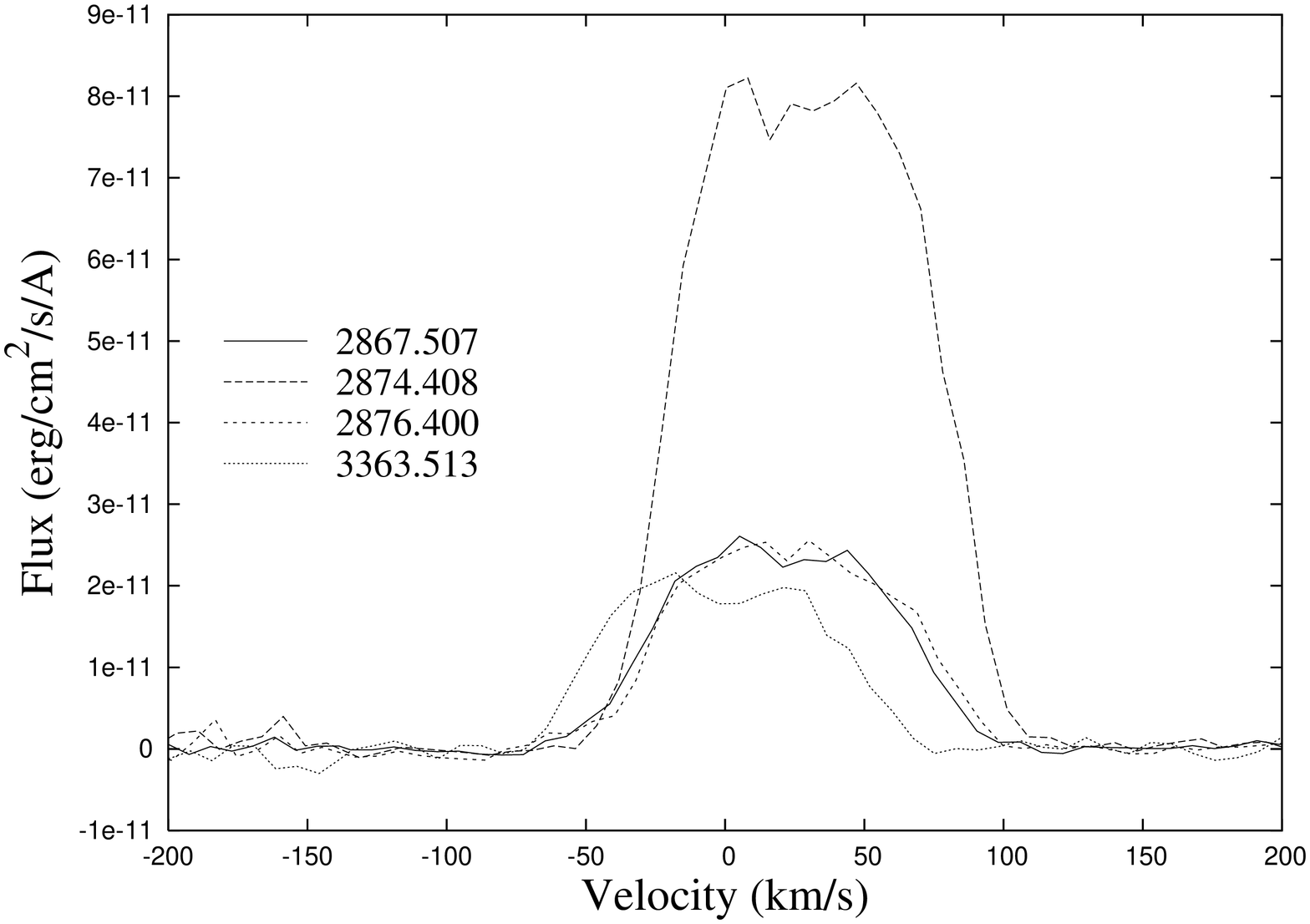,width=5.5cm,angle=-90}

\epsfig{file=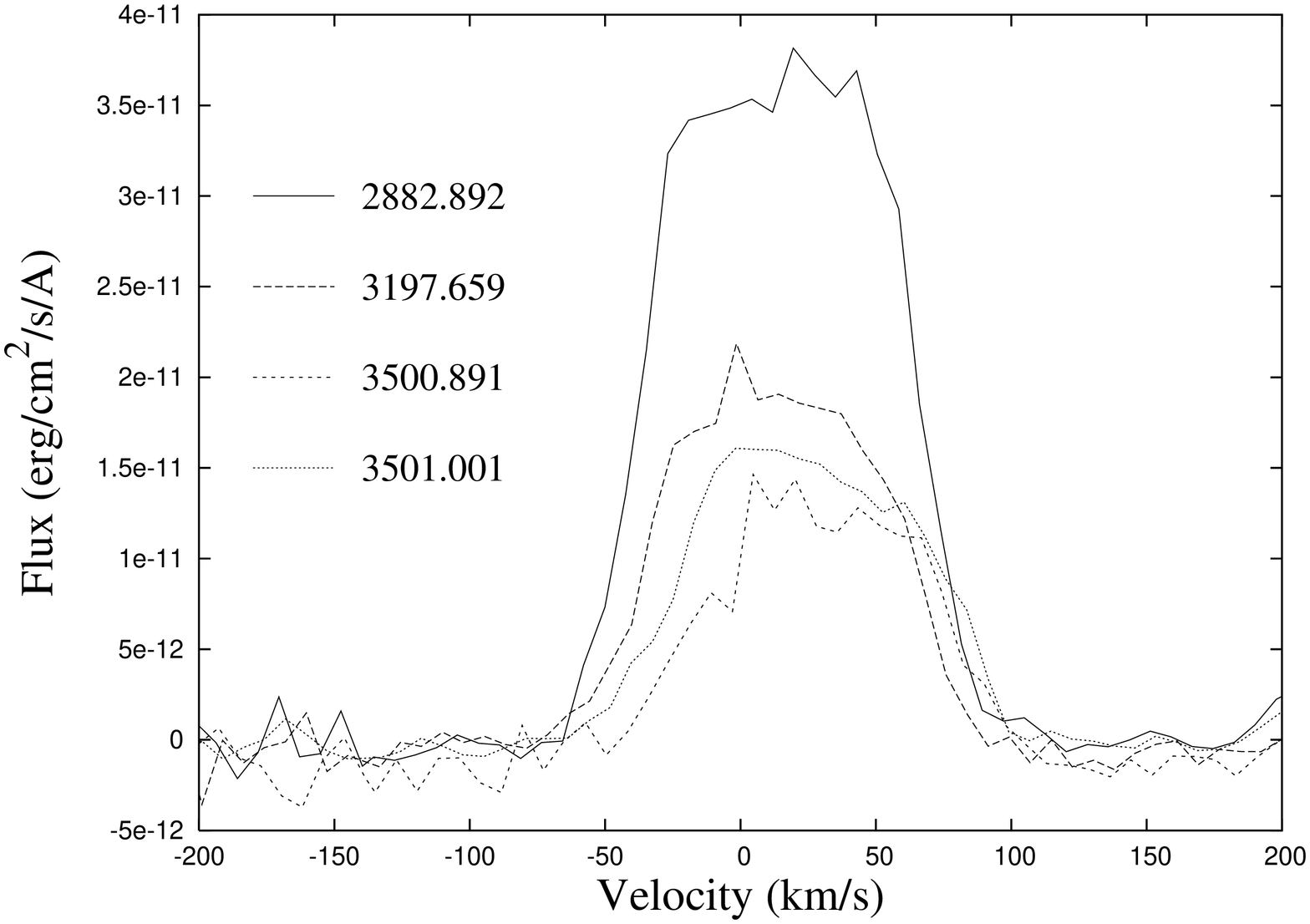,width=5.5cm,angle=-90}
\caption{
The Ly$\alpha$ line of TT Hya (a) at different phases out of eclipse
during a single epoch (top frame); (b) at similar phases close to
secondary minimum but at different epochs (middle frame); and (c) at
different epochs and phases including total eclipse.  The expected
amplitude of the velocity curve of the primary star is about $\pm30$
km\,s$^{-1}$.  The velocities are shifted to the frame of the center of
mass of the system.
}
\label{f13}
\end{figure*}

Another interesting feature can be observed in the new high resolution
HIRES spectra.  The Ly$\alpha$ core is observed in emission and this
emission is very sharp, with a maximum half-width of $\sim 70$
km\,s$^{-1}$.  Figure \ref{f13} shows that the Ly$\alpha$ line is
relatively stable during a single epoch, independent of phase (e.g., epoch
3363; Figure \ref{f13}(a)).  However, Figures \ref{f13}(b), and
\ref{f13}(c) suggest that the $Ly{\alpha}$ emission can be highly
variable.  It can undergo a sudden enhancement, a small wavelength shift,
or display an asymmetry as illustrated in Figure \ref{f13}(b).  The
spectra at epoch 3363 (Fig.  \ref{f13}(a)) are all very similar and
slightly blueshifted relative to the center of mass of the binary, while
those at epochs 2867, 2876, 3500, and 3501 are nearly identical but
slightly redshifted (Fig.  \ref{f13}(b), \ref{f13}(c)).  A flare-like
event or outburst occurred between epochs 2867 and 2874, and was gone by
epoch 2876, within two orbital cycles (Fig.  \ref{f13}(b)).

There are several possible explanations of this Ly$\alpha$ line behavior:
(a) NLTE effects in the atmospheres of the stars; (b) a temperature
inversion in the atmospheres of the stars; (c) emission originating from
extended volumes of circumstellar material; or (d) emission not
originating in the binary system, but in Earth's geocorona or in our solar
system.  Since, the line does not move according to the orbit of either
star we can reject the first two hypotheses.  Since the line is very
sharp, any circumstellar material would have to be very far from the stars
to avoid line broadening by Keplerian motions.  If the circumstellar
structure is in the form of a shell, we can estimate the inner radius of
this structure (the distance from center of mass of binary) to be $\ge 130
R_{\odot}$ by equating the Keplerian velocity with the maximum half-width
of the line, and assuming that the masses of the stars are
$M_{1}+M_{2}=3.40M_{\odot}$.  The maximum velocities observed during the
outburst event (Fig.  \ref{f13}(b)) were $v \le 110$ km\,s$^{-1}$ measured
from the red wing of the line.  Subsequently, the volume of gas involved
in the outburst would have a radius of $R_{max} = v.t \le 190 R_{\odot}$.
However, more data are needed to understand the origin of these outbursts
and hypothesis (d) provides the most likely explanation for the L$\alpha$ 
emission.

\section{Doppler Tomography of the H$\alpha$ Spectra}
\label{s9}

Doppler tomography has been a useful tool for providing images of the
complex accretion structures in the Algols (cf.  \citealt{richards01,
richards04} and references therein).  In this work, we have used this
technique to make images of the accretion structures in the binary, in order 
to examine the quality of the fits between the observed and
synthetic H$\alpha$ spectra, and to demonstrate the effectiveness of the
{\sc{shellspec}} code in modeling selected accretion structures.  The
synthetic H$\alpha$ spectra used in this section were described in
\S3.4 with the disk properties derived from the analysis of the UV
spectra (see \S 4.2 and Table \ref{t6}).  For these calculations, we 
used the newly derived mass and radius of the primary (M = 2.77 $M_\odot$, 
R = 1.99 $R_\odot$; see Table \ref{t5}).  The assumed disk temperature was 
7000 K and the initial gas stream temperature was 8000 K.  We also calculated 
synthetic spectra using the values derived by \citet{vanhamme+wilson93} 
(M = 2.63 $M_\odot$, R = 1.95 $R_\odot$) and found that the two sets of 
synthetic spectra were nearly identical.  As expected, the 
profiles are more sensitive to the temperature and density of the disk 
than to small changes in the stellar mass and radius. 

\begin{figure*}
\figurenum{14}
\epsscale{0.9}
\plotone{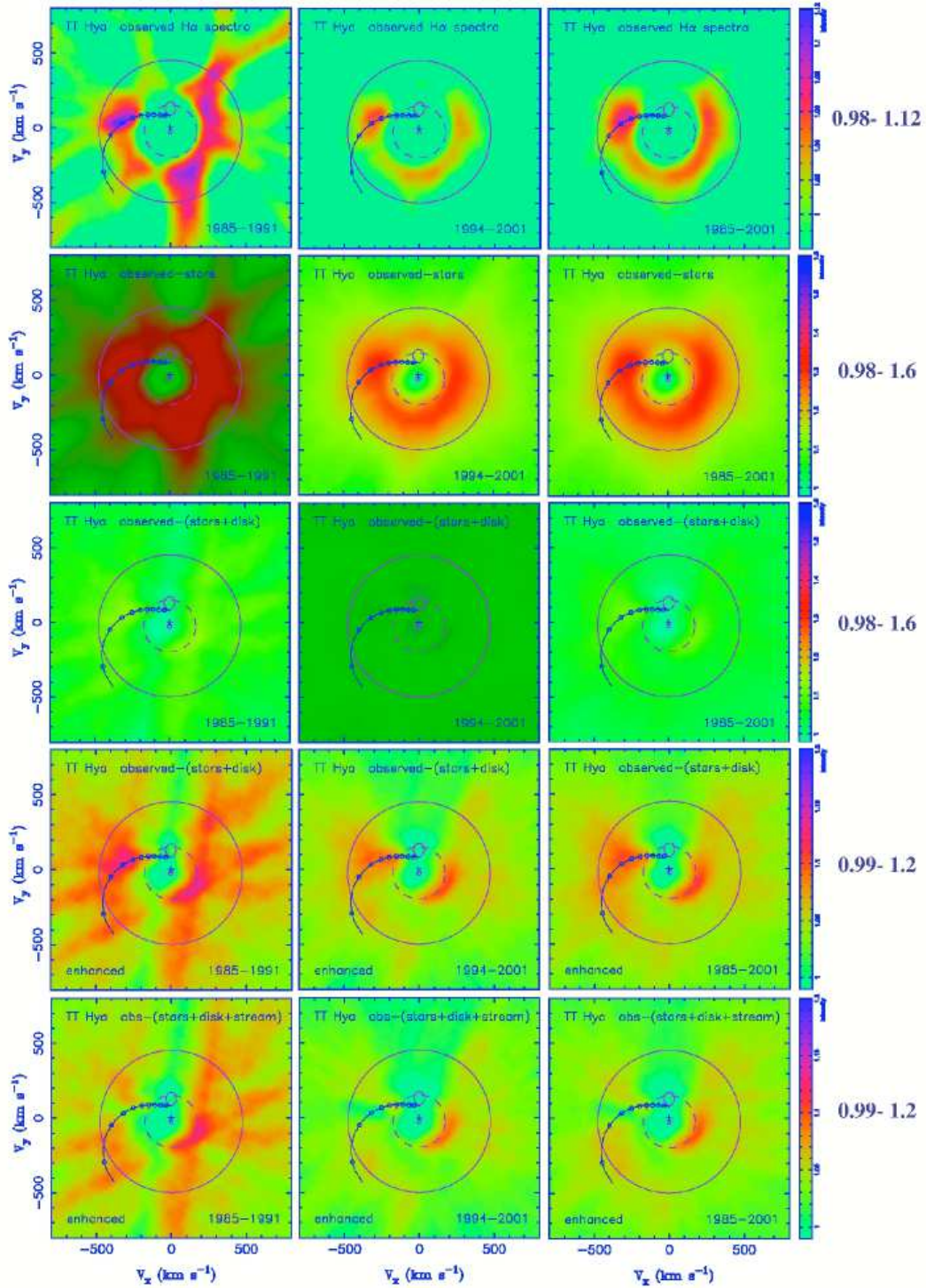}
\caption{
H$\alpha$ Doppler tomograms of TT Hya for (1) 1985--1991 (left column),
(2) 1994--2001 (middle column), and (3) all data from 1985--2001 (right
column).  The different rows show the (a) observed spectra (row 1; Image
intensity, I = 0.98 -- 1.12), (b) observed spectra minus stars (row 2; I =
0.98 -- 1.6), (c) observed spectra minus stars and disk (row 3; I = 0.98
-- 1.6), (d) enhanced version of row 3 (row 4; I = 0.99 -- 1.2), (d)
enhanced version of observed spectra minus stars, disk, and gas stream
(row 5; I = 0.99 -- 1.2).
}
\label{f14}
\end{figure*}

The Doppler tomograms were derived from the 1985--2001 H$\alpha$ data in
several groups:  (a) the observed spectra, (b) the observed spectra minus
the synthetic stellar spectra, (c) the observed spectra minus the
synthetic spectra of the stars and accretion disk, and (d) the observed
spectra minus the synthetic spectra of the stars, disk, and gas stream.
These images were derived for three data sets to illustrate the changes
with time:  (1) 1985 -- 1991, (2) 1994 -- 2001, and (3) the entire data
set from 1985--2001.  The reconstructed images of the H$\alpha$ accretion
sources are shown in Figure \ref{f14}.

Figure \ref{f14} (row 1) shows that the observed H$\alpha$ emission
originated from a part of the disk structure whose outer edge was close to
the Roche surface of the star but whose inner edge did not touch the
surface of the primary.  The tomograms shown in Figure \ref{f14} (row 2)
were based on the difference spectra when only the stars were included in
the synthetic spectrum.  They shows that the accretion disk dominated the
emission profile once the stellar contribution was removed from the
observed spectrum.  When the stars and disk were all modeled in
{\sc{shellspec}}, and difference profiles were calculated, the resulting
image was almost clear (Fig.  \ref{f14}, row 3); however enhancement of
this image showed that two structures remained:  emission along the gas
stream and a portion of the accretion disk (Fig.  \ref{f14}, row 4).
Finally, when the stars, disk, and gas stream were all incorporated into
the synthetic profile, the tomogram showed an almost clear image once
again as in (Fig.  \ref{f14}, row 3), but enhancement of this final image
showed that the gas stream emission had disappeared, and the only
structure that remained arose from the locus of the accretion disk (Fig.
\ref{f14}, row 5).  The main difference between the 1985--1991 (column 1)
and 1994--2001 (column 2) data sets was that the disk emission was
generally stronger during the earlier epoch than in the later epoch.  In
addition, the phase coverage during the earlier epoch was not as good as
during 1994--2001, with only a few spectra between $\phi = 0.4-0.7$ (Fig. 
\ref{f1}), which explains the streaky appearance of the 1985--1991
images.

These Doppler images suggest that the {\sc{shellspec}} code can be used to
sequentially model the stars, disk, and gas stream.  Only a very faint
portion of the disk could not be modeled.  Comparisons with simulated
H$\alpha$ tomograms of TT Hya based on hydrodynamic simulations by
\citet{richards+ratliff98} suggested that the unmodeled region was indeed
part of an asymmetric accretion disk.  This region was not modeled by
{\sc{shellspec}} because the code assumes a circular Keplerian disk structure.
This result provides the first observational confirmation that the disk in
TT Hya is elliptical and asymmetric.  Moreover, the similarity between the
tomogram based on the observed H$\alpha$ spectra only and that based on the observed
minus stellar contribution suggests that the gas in the accretion disk is
optically thin at H$\alpha$, as assumed in the differencing procedure (see
\citealt{richards93,richards04}).  Tomograms of another Algol binary, CX
Dra, suggested that the disk in that system was not optically thin
\citep{richards+koubskyetal00}.

\section{Conclusions}
The main achievements of this work can be summarized as follows:
\begin{itemize}

\item
We have developed a systematic procedure to study the disks in Algol-type
binaries using spectroscopic analysis, synthetic spectra, and tomography.
This procedure can be used to study other Algol disk systems like TT Hya
(e.g., AU Mon).  With minor modifications, this approach can be used to 
study the disks in novae and cataclysmic variables.

\item
New velocity curves and orbital parameters have been derived for TT Hya
from an analysis of H$\alpha$ and UV spectra.  Our study has more than 
quadrupled the previous 27 radial velocity measurements and nearly
tripled their precision.

\item
The derived orbital eccentricity was found to be small, but nonzero ($e =
0.021 \pm 0.003$ based on the best photometric fit).  An eccentric orbit
could lead to enhanced mass transfer near and shortly after periastron,
which might explain the enhanced activity of TT Hya and other Algols at
these phases.

\item
A lower limit of the mass transfer rate was estimated
($2\times10^{-10}M_{\odot}yr^{-1}$).

\item
The calculations of the synthetic H$\alpha$ spectra have been improved
within the {\sc{shellspec}} code compared to those of
\citet{budaj+richards+miller05}.  Specifically, the electron number
density is now calculated within the code from the density, temperature,
and abundances.  The inclusion of the gas stream and the adoption of a
lower surface gravity for the primary star led to an improved match
between the observed and synthetic spectra.

\item
High resolution IUE spectra of TT Hya outside of the eclipses were
compared with synthetic spectra calculated by {\sc{shellspec}}.  We have
demonstrated that the ultraviolet spectrum results from the combined
contributions of the primary star and cooler surrounding disk.

\item
The ultraviolet spectra were used to refine the properties of the
accretion disk.  A reliable disk temperature of about $T=7000$K was found
from comparisons between the observed and synthetic UV spectra of the
Fe\ii\ and Al\iii\ lines in the iron curtain at out-of-eclipse phases.

\item
Radial velocity measurements of the IUE spectra confirmed that they are
decoupled from the radial velocity curve of the primary star.  We propose
that they originate from elliptical stream lines within the accretion
disk.

\item
The comparison between the observed and synthetic UV spectra showed that
the emission lines of higher ionization species (e.g., Al\iii, Si\iv,
C\iv) seen during total eclipse cannot originate in the disk.  These
emission lines could originate from the stream-disk interaction region.

\item
The non-stellar pseudo-continuum during the total eclipse was studied and
quantitatively explained by calculations of synthetic spectra.  This
continuum is most likely due to the thermal and scattered emission in the
lines which becomes unresolved at lower resolution.

\item
Doppler tomography of the observed H$\alpha$ line confirmed the dominant
disk-like structure of the circumstellar material.  Tomography of the
difference spectra after subtracting the synthetic spectrum of the stars
and disk revealed the presence of the gas stream.  This gas stream was
modeled effectively with the {\sc{shellspec}} code once accurate disk and
gas stream parameters were known.

\item
The presence of an asymmetric disk was suggested by the non-circular
nature of the radial velocity measurements of the UV lines and confirmed
by the comparison between the H$\alpha$ tomograms and simulated tomograms
based on hydrodynamic simulations of \citet{richards+ratliff98}.

\end{itemize}

\acknowledgements
We would like to thank Richard Wade, George Pavlov, and Alon Retter for 
many stimulating discussions about this work.  This research was supported 
by a postdoctoral fellowship from Penn State University, NSF-NATO fellowship
DGE-0312144, and NASA ADP grant NNG04GC48G.

\clearpage

\clearpage

\clearpage

\clearpage

\clearpage

\clearpage

\begin{figure*}
\figurenum{15}
\epsscale{1.0}
\plotone{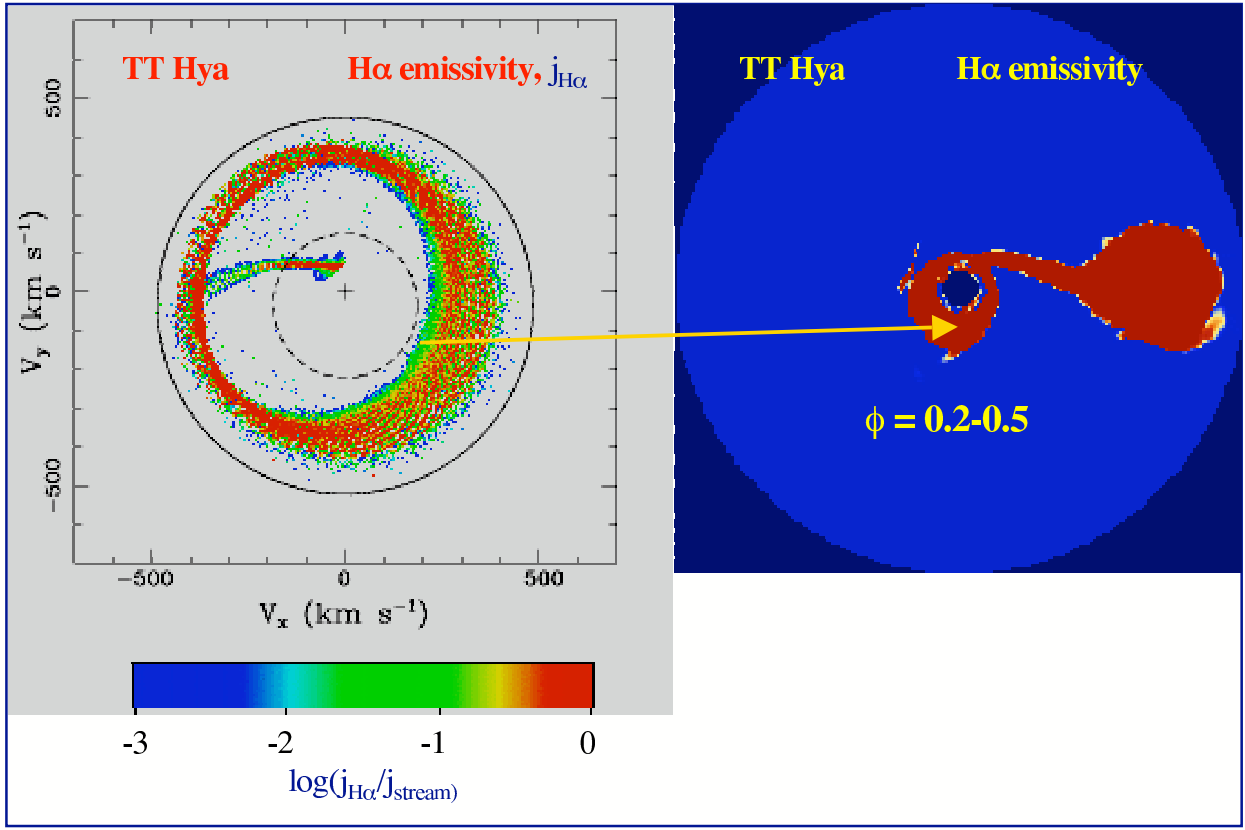}
\caption{
Eccentric disk in TT Hya, with asymmetry near phases 0.2 -- 0.5, based on
hydrodynamic simulations by \citet{richards+ratliff98}.
}
\label{f15}
\end{figure*}
\clearpage

\begin{deluxetable}{llllcl}
\tabletypesize{\scriptsize}
\tablenum{1}
\tablewidth{6.5in}
\tablecaption{Summary of the Optical Observations of TT Hya \label{t1}}
\tablehead{Dates & Observatory & Chip & Resolution  &  Data & Observers}
\startdata
1956 Apr - 1977 Apr & Mt. Wilson/Lick & --     & 11-20 \AA/mm  & 27        & Popper (radial velocities only)\\
1985 Feb - 1991 Mar & KPNO            & TI3  & 0.221 \AA/pix & 57        & Peters \\
1994 Apr - 1994 Apr & KPNO            & T1KA & 0.166 \AA/pix & ~4        & Peters \\
1994 Jun - 1994 Dec & KPNO            & T1KA & 0.166 \AA/pix & 23        & Richards, Albright, Koubsk{\'y} \\
1996 Feb - 1997 Jun & KPNO            & F3KB & 0.105 \AA/pix & 24        & Richards \& Koubsk{\'y}  \\
1999 Jan - 2001 Jan & KPNO            & F3KB & 0.105 \AA/pix & 11        & Peters \\
\enddata
\end{deluxetable}

\clearpage

\begin{deluxetable}{lcccllrrrc}
\tabletypesize{\scriptsize}
\tablenum{2}
\tablewidth{6.5in}
\tablecaption{List of Optical Observations \label{t2}}
\tablehead{UT Date & HJD-2400000 &  Epoch.phase & $I(ab_{c})$\tablenotemark{1}   & $I(em_{b})$   & I($em_{r})$   & $V_r(ab_{c})$ & $V_r(2)$ & (O-C)  & Sp. Phase \\
  &  (days)     &   &     &     &
 & (km\,s$^{-1}$) & (km\,s$^{-1}$) & (km\,s$^{-1}$) &            
}
\startdata
1985 Feb 25 & 46121.8071 & 3092.923 & 0.655& 0.960& 1.235& -10.8 &  -59.5 &  3.2 & 0.394 \\
1985 Feb 25 & 46121.9368 & 3092.942 & 0.510& 0.920& 1.300&  -    &   ---- & ---- & ----  \\
1985 Feb 26 & 46122.7729 & 3093.062 & 0.380& 1.185& 1.025&  36.9 &   42.7 & -4.3 & 0.533 \\
1985 Feb 26 & 46122.8474 & 3093.073 & 0.430& 1.240& 1.105&  42.0 &   54.1 & -1.0 & 0.543 \\
1985 Feb 26 & 46122.9285 & 3093.085 & 0.500& 1.260& 1.170&  46.8 &   62.2 & -1.4 & 0.555 \\
1985 Mar 24 & 46148.8856 & 3096.818 & 0.800& 1.145& 1.180&  69.9 & -120.1 &  2.8 & 0.288 \\
1985 Mar 25 & 46149.8922 & 3096.962 & 0.480& 0.960& 1.480&  43.8 &   ---- & ---- & ----  \\
1985 Mar 28 & 46152.7334 & 3097.371 & 0.750& 1.170& 1.170&  -    &   ---- & ---- & ----  \\
1987 Apr 16 & 46901.7567 & 3205.091 & 0.640& 1.340& 1.220&  61.2 &   69.4 &  1.9 & 0.560 \\
1987 Apr 16 & 46901.8083 & 3205.098 & 0.650& 1.320& 1.240&  60.5 &   77.0 &  4.3 & 0.568 \\
1987 Apr 16 & 46901.8374 & 3205.103 & 0.660& 1.340& 1.240&  53.8 &   80.6 &  5.1 & 0.572 \\
1987 Apr 17 & 46902.7817 & 3205.238 & 0.720& 1.215& 1.190&  -2.8 &  136.6 &  4.1 & 0.708 \\
1987 Apr 18 & 46903.8010 & 3205.385 & 0.665& 1.220& 1.260&  16.3 &   89.2 & -4.2 & 0.854 \\
1987 Apr 20 & 46905.7692 & 3205.668 & 0.700& 1.350& 1.240&  33.3 & -120.6 & -1.0 & 0.137 \\
1987 May 02 & 46917.7436 & 3207.390 & 0.610& 1.180& 1.180&   1.8 &   90.8 &  0.6 & 0.859 \\
1987 May 03 & 46918.7052 & 3207.528 & 0.520& 1.240& 1.360& -24.3 &  -26.7 & -4.4 & 0.998 \\
1987 May 04 & 46919.7251 & 3207.675 & 0.665& 1.240& 1.140&  39.9 & -124.2 & -1.8 & 0.144 \\
1987 May 05 & 46920.7267 & 3207.819 & 0.760& 1.150& 1.180&  52.6 & -125.2 & -2.4 & 0.288 \\
1987 May 06 & 46921.7241 & 3207.963 & 0.400& 0.915& 1.400&  -9.2 &  -32.6 &  0.8 & 0.432 \\
1987 May 06 & 46921.7913 & 3207.972 & 0.470& 1.070& 1.740&  -6.1 &  -24.8 &  0.9 & 0.442 \\
1987 May 06 & 46921.8143 & 3207.976 & 0.490& 1.160& 1.890&  -7.5 &  -22.6 &  0.4 & 0.445 \\
1988 May 21 & 47302.6684 & 3262.748 & 0.670& 1.145& 1.170&  44.0 & -135.1 &  1.2 & 0.217 \\
1988 May 21 & 47302.7044 & 3262.753 & 0.660& 1.140& 1.170&  51.0 & -133.5 &  2.7 & 0.222 \\
1988 May 22 & 47303.6591 & 3262.890 & 0.680& 1.190& 1.270&  14.8 &  -84.5 &  2.1 & 0.359 \\
1988 May 22 & 47303.7218 & 3262.899 & 0.675& 1.155& 1.310&   6.2 &  -76.0 &  4.7 & 0.368 \\
1988 May 22 & 47303.7523 & 3262.904 & 0.660& 1.135& 1.320& -12.4 &  -73.0 &  4.7 & 0.372 \\
1988 May 23 & 47304.6429 & 3263.032 & 0.430& 1.650& 1.110&  39.8 &   24.7 &  2.7 & 0.500 \\
1988 May 23 & 47304.6701 & 3263.036 & 0.390& 1.510& 1.000&  40.9 &   26.2 &  1.0 & 0.504 \\
1988 May 23 & 47304.7134 & 3263.042 & 0.345& 1.370& 0.920&  43.0 &   28.7 & -1.2 & 0.511 \\
1988 May 23 & 47304.7405 & 3263.046 & 0.330& 1.300& 0.870&  45.3 &   32.6 & -0.5 & 0.515 \\
1988 May 24 & 47305.6447 & 3263.176 & 0.700& 1.200& 1.220&  14.0 &  118.2 &  2.5 & 0.645 \\
1988 May 24 & 47305.7153 & 3263.186 & 0.710& 1.210& 1.210&   8.6 &  120.0 &  0.4 & 0.655 \\
1988 May 25 & 47306.6893 & 3263.326 & 0.690& 1.130& 1.210& -26.4 &  122.2 & -0.2 & 0.795 \\
1988 Nov 04 & 47470.0043 & 3286.813 & 0.730& 1.110& 1.110&  78.4 & -121.6 &  3.6 & 0.282 \\
1988 Nov 05 & 47470.9921 & 3286.955 & 0.410& 0.850& 1.300&  -1.9 &  -41.4 & -1.4 & 0.424 \\
1988 Nov 06 & 47471.9933 & 3287.099 & 0.560& 1.220& 1.180&  51.6 &   62.9 & -9.6 & 0.568 \\
1989 Feb 03 & 47560.9107 & 3299.887 & 0.790& 1.220& 1.300&  26.9 &  -86.8 &  2.3 & 0.355 \\
1989 Feb 03 & 47560.9919 & 3299.898 & 0.770& 1.170& 1.310&  49.8 &  -77.0 &  4.5 & 0.367 \\
1989 Feb 03 & 47561.0249 & 3299.903 & 0.775& 1.170& 1.340&  52.3 &  -76.3 &  2.1 & 0.371 \\
1989 Feb 04 & 47561.7990 & 3300.014 & 0.680& 2.440& 1.720&  26.0 &   13.8 &  6.1 & 0.483 \\
1989 Feb 04 & 47561.8334 & 3300.019 & 0.660& 2.450& 1.570&  32.0 &   17.8 &  6.1 & 0.488 \\
1989 Feb 04 & 47561.9813 & 3300.041 & 0.380& 1.420& 0.980&  35.7 &   32.3 &  3.5 & 0.509 \\
1989 Feb 04 & 47562.0055 & 3300.044 & 0.350& 1.350& 0.940&  37.6 &   34.4 &  2.8 & 0.512 \\
1989 Apr 20 & 47636.7293 & 3310.790 & 0.800& 1.220& 1.230&  46.2 & -136.4 & -4.8 & 0.259 \\
1989 Apr 20 & 47636.7744 & 3310.797 & 0.780& 1.195& 1.235&  48.7 & -131.8 & -1.7 & 0.265 \\
1989 Apr 21 & 47637.7765 & 3310.941 & 0.550& 0.930& 1.330&  -8.0 &  -39.6 & 11.4 & 0.409 \\
1989 Apr 23 & 47639.7354 & 3311.223 & 0.770& 1.250& 1.230&  -3.4 &  136.6 &  6.7 & 0.691 \\
1989 Apr 23 & 47640.7299 & 3311.366 & 0.665& 1.200& 1.320&  12.5 &   99.1 & -6.0 & 0.834 \\
1990 Feb 17 & 47939.8551 & 3354.384 & 0.645& 1.210& 1.280&  -2.8 &   91.5 & -3.2 & 0.852 \\
1990 Feb 18 & 47940.8822 & 3354.532 & 0.380& 1.160& 1.370& -27.7 &  -31.8 & -7.7 & 0.000 \\
1991 Feb 26 & 48313.8742 & 3408.173 & 0.600& 1.180& 1.180&   1.2 &  119.1 &  5.1 & 0.641 \\
1991 Feb 27 & 48314.8265 & 3408.310 & 0.760& 1.160& 1.210& -33.3 &  125.1 & -2.4 & 0.778 \\
1991 Mar 04 & 48319.8424 & 3409.031 & 0.510& 1.720& 1.120&  23.0 &   13.4 & -7.3 & 0.499 \\
1991 Mar 04 & 48319.8759 & 3409.036 & 0.460& 1.580& 1.040&  24.3 &   17.4 & -7.2 & 0.504 \\
1991 Mar 04 & 48319.9089 & 3409.041 & 0.420& 1.450& 0.960&  24.8 &   19.2 & -9.2 & 0.509 \\
1991 Mar 04 & 48319.9510 & 3409.047 & 0.400& 1.370& 0.900&  29.6 &   23.8 & -9.4 & 0.515 \\
1991 Mar 05 & 48320.8750 & 3409.180 & 0.630& 1.240& 1.200& -11.2 &  113.2 & -3.6 & 0.648 \\
1994 Apr 01 & 49443.7821 & 3570.670 & 0.720& 1.460& 1.370&  27.9 & -123.4 & -4.5 & 0.136 \\
1994 Apr 02 & 49444.7740 & 3570.812 & 1.065& 1.275& 1.260&  74.8 & -127.9 & -1.7 & 0.279 \\
1994 Apr 03 & 49445.7918 & 3570.959 & 0.550& 0.970& 1.480&  -8.5 &  -36.7 &  2.2 & 0.425 \\
1994 Apr 05 & 49447.8002 & 3571.248 & 0.750& 1.350& 1.230&   2.0 &  133.7 &  0.6 & 0.714 \\
1994 Jun 02 & 49505.6359 & 3579.565 & 0.57 & 1.07 & 1.24 &  -4.9 &  -44.3 & -2.6 & 0.031 \\
1994 Jun 03 & 49506.6465 & 3579.711 & 0.71 & 1.06 & 1.15 &  48.4 & -121.8 &  1.0 & 0.177 \\
1994 Jun 04 & 49507.6476 & 3579.854 & 0.79 & 1.10 & 1.17 &   9.8 & -104.2 & -4.6 & 0.321 \\
1994 Jun 05 & 49508.6534 & 3579.999 & 0.82 & 1.73 & 1.83 &  -2.6 &    3.3 &  0.7 & 0.465 \\
1994 Jun 05 & 49508.6795 & 3580.003 & 0.82 & 1.84 & 1.76 &   0.3 &    6.1 &  0.4 & 0.469 \\
1994 Jun 06 & 49509.6547 & 3580.143 & 0.81 & 1.21 & 1.14 &   6.5 &  111.3 &  3.5 & 0.609 \\
1994 Jun 07 & 49510.6480 & 3580.286 & 0.76 & 1.12 & 1.09 & -24.2 &  143.6 &  1.9 & 0.752 \\
1994 Jun 08 & 49511.6477 & 3580.430 & 0.54 & 1.20 & 1.11 &  20.1 &   73.7 & -0.1 & 0.896 \\
1994 Dec 02 & 49688.9327 & 3605.926 & 0.57 & 0.99 & 1.25 & -13.3 &  -55.0 & -0.0 & 0.392 \\
1994 Dec 02 & 49688.9703 & 3605.931 & 0.54 & 0.95 & 1.26 &  -8.3 &  -51.5 & -0.4 & 0.397 \\
1994 Dec 02 & 49689.0150 & 3605.938 & 0.50 & 0.92 & 1.25 &  -3.0 &  -46.1 &  0.1 & 0.403 \\
1994 Dec 02 & 49689.0331 & 3605.940 & 0.48 & 0.91 & 1.26 &  -1.9 &  -45.1 & -0.7 & 0.406 \\
1994 Dec 02 & 49689.0584 & 3605.944 & 0.45 & 0.89 & 1.24 &  3.00 &  -42.4 & -0.8 & 0.410 \\
1994 Dec 03 & 49689.9092 & 3606.066 & 0.34 & 1.26 & 0.95 &  36.3 &   54.6 & -1.4 & 0.532 \\
1994 Dec 03 & 49689.9404 & 3606.071 & 0.36 & 1.24 & 0.98 &  38.6 &   59.1 & -0.3 & 0.537 \\
1994 Dec 03 & 49689.9815 & 3606.077 & 0.38 & 1.22 & 1.01 &  43.6 &   62.7 & -1.1 & 0.542 \\
1994 Dec 03 & 49690.0391 & 3606.085 & 0.41 & 1.21 & 1.07 &  44.7 &   69.1 & -0.8 & 0.551 \\
1994 Dec 04 & 49690.9599 & 3606.217 & 0.68 & 1.14 & 1.13 & -24.2 &  139.4 &  1.8 & 0.683 \\
1994 Dec 04 & 49691.0158 & 3606.225 & 0.67 & 1.13 & 1.13 & -30.1 &  140.4 &  1.1 & 0.691 \\
1994 Dec 04 & 49691.0562 & 3606.231 & 0.67 & 1.13 & 1.10 & -27.6 &  143.1 &  2.8 & 0.697 \\
1994 Dec 08 & 49694.9391 & 3606.790 & 0.78 & 1.17 & 1.17 &  50.8 & -122.7 &  0.4 & 0.255 \\
1994 Dec 08 & 49694.9881 & 3606.797 & 0.79 & 1.17 & 1.17 &  55.2 & -120.2 &  1.2 & 0.262 \\
1994 Dec 08 & 49695.0535 & 3606.806 & 0.80 & 1.18 & 1.18 &  63.7 & -120.0 & -0.9 & 0.272 \\
1996 Feb 16 & 50129.7633 & 3669.323 & 0.75 & 1.12 & 1.13 & -27.8 &  132.3 & -1.3 & 0.789 \\
1996 Feb 16 & 50129.8095 & 3669.330 & 0.73 & 1.12 & 1.14 & -29.6 &  129.4 & -2.0 & 0.795 \\
1996 Feb 16 & 50129.8569 & 3669.337 & 0.73 & 1.12 & 1.17 & -32.5 &  127.6 & -1.4 & 0.802 \\
1996 Feb 17 & 50130.7779 & 3669.469 & 0.47 & 1.26 & 1.13 &   0.5 &   42.6 &  0.3 & 0.935 \\
1996 Feb 17 & 50130.8203 & 3669.475 & 0.45 & 1.25 & 1.14 &   1.2 &   39.9 &  2.8 & 0.941 \\
1996 Feb 17 & 50130.8617 & 3669.481 & 0.42 & 1.26 & 1.16 &  -0.6 &   34.4 &  2.5 & 0.947 \\
1996 Feb 17 & 50130.9299 & 3669.491 & 0.39 & 1.28 & 1.19 &  -1.5 &   26.3 &  3.0 & 0.956 \\
1996 Feb 17 & 50130.9601 & 3669.496 & 0.38 & 1.26 & 1.21 &  -2.0 &   24.3 &  4.8 & 0.961 \\
1996 Feb 17 & 50130.9929 & 3669.500 & 0.38 & 1.25 & 1.22 &  -2.4 &   17.1 &  1.7 & 0.966 \\
1996 Feb 18 & 50131.7747 & 3669.613 & 0.67 & 1.32 & 1.23 &  32.4 &  -80.7 & -3.5 & 0.078 \\
1996 Feb 18 & 50131.8171 & 3669.619 & 0.67 & 1.31 & 1.22 &  36.5 &  -83.4 & -2.2 & 0.084 \\
1996 Feb 18 & 50131.8638 & 3669.625 & 0.68 & 1.30 & 1.23 &  35.6 &  -86.1 & -0.5 & 0.091 \\
1996 Feb 18 & 50131.9239 & 3669.634 & 0.68 & 1.31 & 1.22 &  34.2 &  -90.5 &  0.3 & 0.099 \\
1996 Feb 18 & 50131.9604 & 3669.639 & 0.68 & 1.31 & 1.22 &  38.0 &  -94.2 & -0.1 & 0.105 \\
1996 Feb 19 & 50132.7798 & 3669.757 & 0.75 & 1.11 & 1.17 &  64.9 & -125.4 &  1.7 & 0.222 \\
1996 Feb 19 & 50132.8731 & 3669.771 & 0.73 & 1.13 & 1.15 &  ---- &  ----  & ---- & ----  \\
1996 Feb 19 & 50132.9687 & 3669.784 & 0.74 & 1.09 & 1.17 &  ---- &  ----  & ---- & ----  \\
1997 May 31 & 50599.6386 & 3736.898 & 0.72 & 1.17 & 1.26 &  30.5 &  -77.0 & -2.0 & 0.363 \\
1997 Jun 01 & 50600.6198 & 3737.039 & 0.43 & 1.57 & 1.02 &  31.5 &   33.4 & -0.4 & 0.504 \\
1997 Jun 02 & 50601.6166 & 3737.187 & 0.59 & 1.17 & 1.16 & -15.9 &  128.5 &  2.5 & 0.647 \\
1997 Jun 03 & 50602.6167 & 3737.326 & 0.69 & 1.13 & 1.17 & -23.2 &  133.0 &  0.0 & 0.791 \\
1997 Jun 04 & 50603.6209 & 3737.471 & 0.42 & 1.22 & 1.17 &   5.2 &   39.9 & -1.5 & 0.935 \\
1997 Jun 05 & 50604.6189 & 3737.614 & 0.62 & 1.32 & 1.18 &  -7.3 &  -79.8 & -1.9 & 0.079 \\
1997 Jun 07 & 50606.6175 & 3737.902 & 0.57 & 1.07 & 1.16 & -15.4 &  ----  & ---- & ----  \\
1999 Jan 13 & 51191.9724 & 3822.084 & 0.440& 1.270& 1.100&  47.7 &   57.3 & -1.2 & 0.548 \\
1999 Jan 14 & 51192.9756 & 3822.228 & 0.640& 1.190& 1.180&  -8.1 &  130.1 & -0.0 & 0.692 \\
1999 Jan 15 & 51194.0161 & 3822.378 & 0.690& 1.160& 1.230&   3.3 &  102.2 &  1.3 & 0.842 \\
1999 Jan 17 & 51196.0222 & 3822.666 & 0.610& 1.240& 1.110&  38.3 & -116.1 &  0.3 & 0.130 \\
1999 Jan 18 & 51196.9780 & 3822.804 & 0.740& 1.100& 1.140&  56.9 & -126.4 &  3.0 & 0.268 \\
1999 Jan 19 & 51197.9733 & 3822.947 & 0.470& 0.870& 1.240& -15.6 &  -50.5 & -0.6 & 0.411 \\
1999 Jan 19 & 51198.0140 & 3822.953 & 0.420& 0.840& 1.240& -12.4 &  -45.6 & -0.1 & 0.417 \\
1999 Jan 19 & 51198.0353 & 3822.956 & 0.400& 0.825& 1.230& -11.8 &  -43.3 & -0.2 & 0.420 \\
1999 Nov 21 & 51504.0479 & 3866.965 & 0.400& 0.880& 1.400&  -8.1 &  -34.6 &  1.7 & 0.428 \\
2001 Jan 05 & 51915.0339 & 3926.070 & 0.380& 1.390& 1.000&  46.2 &   49.0 &  1.3 & 0.533 \\
2001 Jan 07 & 51916.9961 & 3926.352 & 0.810& 1.220& 1.270&  23.8 &  110.6 & -3.5 & 0.815 \\
\enddata
\tablenotetext{1}{$I(ab_{c})$ is the normalized flux (depth) of the central absorption of H$\alpha$; 
$I(em_{b})$ is the strength of the blue emission peak; 
$I(em_{r})$ is the strength of the red 
emission peak; 
$V_r(ab_{c})$ is the radial velocity of the central absorption (km\,s$^{-1}$); 
$V_r(2)$ is the radial velocity of the secondary (km\,s$^{-1}$);
O-C is the (O-C) for  the radial velocities of the secondary (km\,s$^{-1}$).; and 
Sp. phase is the spectroscopic phase. }
\end{deluxetable}

\clearpage

\begin{deluxetable}{lccllc}
\tabletypesize{\normalsize}
\tablenum{3}
\tablewidth{6.5in}
\tablecaption{High Resolution IUE Observations \label{t3}}
\tablehead{UT Date & HJD-2400000 & Epoch.phase & Region, No. & Observers & $V_r$ \\ 
& & & & & (km\,s$^{-1}$)}
\startdata
1980 Dec 28 & 2444602.3014 & 2874.397 & LWP 10176HL & Plavec & 52.2  \\
1980 Dec 29 & 2444602.5507 & 2874.433 & LWP 17802HL & Peters & 65.4  \\
1987 Feb 23 & 2446850.1432 & 3197.668 & LWP 24387HL & Eaton  & 69.3  \\
1990 Apr 23 & 2448005.3561 & 3363.804 & LWP 24393HL & Eaton  &  ---  \\
1992 Dec 02 & 2448958.6272 & 3500.898 & LWR 09595HL & Kondo  & -20.7 \\
1992 Dec 02 & 2448959.4625 & 3501.018 & LWR 09597HL & Kondo  & -23.2 \\
\hline
1980 Nov 10 & 2444554.3882 & 2867.507 & SWP 10585HL & Kondo  &  7.7  \\
1980 Dec 28 & 2444602.3738 & 2874.408 & SWP 10912HL & Kondo  & -18.9 \\
1981 Jan 11 & 2444616.2290 & 2876.400 & SWP 11024HL & Peters & -25.4 \\
1981 Feb 25 & 2444661.3663 & 2882.892 & SWP 13361HL & Peters &  63.6 \\
1987 Feb 23 & 2446850.0810 & 3197.659 & SWP 30367HL & Plavec &  61.9 \\
1990 Apr 19 & 2448001.4831 & 3363.247 & SWP 38633HL & Peters &  18.1 \\
1990 Apr 21 & 2448003.3308 & 3363.513 & SWP 38643HL & Peters &  11.4 \\
1990 Apr 22 & 2448004.2864 & 3363.650 & SWP 38651HL & Peters &  47.7 \\
1990 Apr 23 & 2448005.2903 & 3363.795 & SWP 38665HL & Peters &  73.2 \\
1992 Dec 02 & 2448958.5819 & 3500.891 & SWP 46386HL & Eaton  & 70.2  \\
1992 Dec 02 & 2448959.3446 & 3501.001 & SWP 46392HL & Eaton  &  ---  \\
\enddata
\end{deluxetable}

\clearpage

\begin{deluxetable}{lllll}
\tabletypesize{\scriptsize}
\tablenum{4}
\tablewidth{6.5in}
\tablecaption{Orbital Elements of TT Hya \label{t4}}
\tablehead{Property & 1994--2001 data     & 1985--2001 data     & 1956--2001 data\tablenotemark{2} & 1956--2001 data\tablenotemark{2}   \\
                    & (Period fixed\tablenotemark{1}) & (Period fixed\tablenotemark{1}) & (Period fixed\tablenotemark{1}) & (Period not fixed)
}
\startdata
Systemic velocity, $V_{0}$ (km\,s$^{-1}$)          & $+8.43\pm0.27$      & $+8.56 \pm 0.28$    & $+8.43\pm0.27$      & $+8.34\pm0.26$       \\   
Velocity semiamplitude, $K_{2}$ (km\,s$^{-1}$)    & $135.76\pm0.40$     & $135.55\pm0.41$     & $135.12\pm0.40$     & $135.13\pm0.37$    \\ 
Eccentricity, $e$                          & $0.0214\pm0.0028$   & $0.0217\pm 0.0029$  & $0.0204\pm0.0029$   & $0.0211\pm0.0027$ \\   
Longitude of periastron, $\omega$ ($^\circ$)       & $99.9\pm 7.5$       & $100.7\pm7.4 $      & $103.4\pm7.9$       & $101.6\pm7.2$      \\    
Epoch of periastron, $T_{0}$ (days)            & $2450006.09\pm0.14$ & $2450006.07\pm0.15$ & $2450006.14\pm0.15$ & $2450006.10\pm0.14$ \\   
Orbital period, $P$ (days)                         & $6.95342913$        & $6.95342913$        & $6.95342913$        & $6.953484\pm0.000012$ \\   
\hline
Mass function, $f(m)$                              & $1.806\pm0.016$     & $1.797\pm0.016$     & $1.781\pm0.016$     & $1.781\pm0.015$   \\
Semi-major axis, $a_{2}\sin i$ ($10^7$ km)        & $1.2978\pm0.0038$   & $1.2957\pm0.0039$   & $1.2917\pm0.0038$   & $1.2917\pm0.0036$\\
Standard deviation of fit, $\sigma$ (km\,s$^{-1}$) & $2.0$          	 & $2.9$               & $3.2$               & $3.0$ \\
Optimum fit                                        & Best photometry &                    &                     & Best spectroscopy  \\
\enddata
\tablenotetext{1}{Photometric period from \citet{kulkarni+abhyankar80}.}
\tablenotetext{2}{Includes data from \citet{popper89} (see Figure \ref{f2}).}
\end{deluxetable}

\clearpage

\begin{deluxetable}{lllll}
\tabletypesize{\normalsize}
\tablenum{5}
\tablewidth{5in}
\tablecaption{The Physical Properties of TT Hya \label{t5}}
\tablehead{Property   & VW\tablenotemark{1} & This paper}
\startdata
Mass ratio, q                      & 0.2261~~  & 0.2261~~ \\
Inclination, i ($^{\circ}$)        & 82.84~~ & 82.84~~ \\
Semi-major axis, $a$ ($R_{\odot}$)  & 22.63~~ & 23.04~~ \\
Mass, $M_{1}$ ($M_{\odot}) $       & ~2.63~~ & ~2.77~~ \\
Mass, $M_{2}$ ($M_{\odot}$)        & ~0.59~~ & ~0.63~~ \\
Radius, $R_{1}$ ($R_{\odot}$)      & ~1.95~~ & ~1.99~~ \\
Radius, $R_{2}$ ($R_{\odot}$)      & ~5.87~~ & ~5.98~~ \\
\enddata
\tablenotetext{1}{These values are from Table 4b of 
\citet{vanhamme+wilson93}.} 
\end{deluxetable}

\clearpage

\begin{deluxetable}{lll}
\tabletypesize{\normalsize}
\tablenum{6}
\tablewidth{5.5in}
\tablecaption{Adopted and Derived Properties of TT Hya from IUE-SWP Spectra\label{t6}}
\tablehead{Primary:&                & Source\tablenotemark{1}}
\startdata
~~Mass, $M$       & $2.77 M_{\odot}$  & MBRKP\\   
~~Radius, $R$       & $1.99 R_{\odot}$ & MBRKP\\   
~~Surface gravity, $\log g$  & 4.                & MBRKP \\ 
~~Temperature, $T$       & 10000 K            & MBRKP\\
~~Rotational velocity, $v\sin i$ & 168 km\,s$^{-1}$	& E\\   
~~Limb darkening coeff., u         & $0.5$           	& VW\\  
\hline
Disk:&                   &\\
~~Inclination, $i$        	& $82.84^{\circ}$	& VW\\
~~Vertical half thickness, $\alpha$   	& $2 R_{\odot}$  	& BRM\\
~~Inner radius, $R_{in}$   	& $2 R_{\odot}$  	& BRM\\
~~Outer radius, $R_{out}$  	& $10 R_{\odot}$	& BRM\\
~~Density, $\rho(R_{in})$  & $5\times 10^{-14}$ g\,cm$^{-3}$ & MBRKP, BRM  \\
~~Temperature, $T$             & 7000 K              & MBRKP\\
~~Density exponent\tablenotemark{2}, $\eta$          & $-1$                & BRM\\
~~Microturbulence, $v_{trb}$               & 30 km\,s$^{-1}$     & BRM\\
\enddata
\tabletypesize{\scriptsize}
\tablenotetext{1}{Here, VW - \citet{vanhamme+wilson93},  E - \citet{etzel88}, 
BRM - \citet{budaj+richards+miller05}, and MBRKP - this work.}
\tablenotetext{2}{Also, $\eta$ is the exponent of the power law in the radial 
density profile (see BRM).}
\end{deluxetable}

\clearpage
\end{document}